\documentclass[letterpaper,twocolumn,10pt]{article}
\usepackage{usenix}
\usepackage{tikz}
\usepackage{amsmath}
\usepackage{xspace}
\usepackage{subcaption}
\usepackage{bigstrut}
\usepackage{booktabs}
\usepackage{array} 
\usepackage{multirow}
\usepackage{enumitem}
\usepackage{color}
\usepackage{xcolor}
\usepackage{xurl} 
\usepackage{url}
\usepackage{hyperref}
\usepackage{arydshln} 
\usepackage{graphicx}
\usepackage{afterpage}
\usepackage{pdflscape}
\usepackage{subcaption}
\usepackage{graphics}
\usepackage{amssymb}
\hypersetup{
    colorlinks=true,
    linkcolor=blue,
    filecolor=magenta,      
    urlcolor=blue,
    pdfpagemode=FullScreen,
    citecolor=blue
    }

\newcommand{\point}[1]{\par\smallskip\noindent\textbf{#1.}}
\newcommand{\empirical}[1]{#1}

\begin{document}

\date{}

\title{Unaligned Incentives: Pricing Attacks Against Blockchain Rollups}

\author{
{\rm Stefanos Chaliasos}\\
Imperial College London\\
zkSecurity
\and
{\rm Conner Swann}\\
Prooflab
\and
{\rm Sina Pilehchiha}\\
Hyperlane
\and
{\rm Nicolas Mohnblatt}\\
Bain Capital Crypto\\
zkSecurity
\and
{\rm Benjamin Livshits}\\
Imperial College London
\and
{\rm Assimakis Kattis}\\
} 

\maketitle

\begin{abstract}
Rollups have become the \textit{de facto} scalability solution for Ethereum, securing more than~\$55B in assets. They achieve scale by executing transactions on a Layer~2 (L2) ledger, while periodically posting data and finalizing state on the Layer~1 (L1), either optimistically or via validity proofs. Their fees must simultaneously reflect the pricing of three resources: L2 costs (e.g., execution/storage), L1 data‐availability (DA), and underlying L1 gas costs for batch settlement and proof verification. In this work, we identify critical \emph{mis-pricings} in existing rollup transaction fee mechanisms~(TFMs) that allow for two powerful attacks. Firstly, an adversary can \emph{saturate} the L2's DA batch capacity with compute-light data-heavy transactions, forcing low-gas transaction batches that enable both L2 Denial of Service (DoS) attacks, and finality-delay attacks. Secondly, by crafting \emph{prover killer} transactions that maximize proving cycles relative to the gas charges, an adversary can effectively \emph{stall} proof generation, delaying finality by hours and inflicting prover-side economic losses to the rollup at a minimal cost.

We analyze the above attack vectors across the major Ethereum rollups, quantifying adversarial costs and protocol losses. We find that the first attack enables periodic DoS on rollups, lasting up to~$30$ minutes, at a cost below~$2$ ETH for most rollups. Moreover, we identify three rollups that are exposed to indefinite DoS at a cost of approximately~$0.8$ to $2.7$ ETH per hour. The attack can be further modified to increase finalization delays by a factor of about~$1.45\times$ to~$2.73\times$, compared to direct L1 blob-stuffing, depending on the rollup's parameters. Furthermore, we find that the prover killer attack induces a finalization latency increase of about~$94\times$. Finally, we propose comprehensive mitigations to prevent these attacks and suggest how some practical uses of multi-dimensional rollup TFMs can rectify the identified mis-pricing attacks.
\end{abstract}

\section{Introduction}
\label{sec:intro}

Scalability has been a central challenge for Layer~1 (L1) blockchains, such as Ethereum~\cite{wood2014ethereum} and Bitcoin~\cite{nakamoto2008bitcoin}. Rollups have recently emerged as the leading Layer~2~(L2) scaling solution for Ethereum, executing transactions off-chain, while periodically posting compressed data and finalizing transactions on the L1 through optimistic or validity proofs. Today, rollups secure over~\$55 billion in assets and sustain more than~250 transactions per second on average, an order of magnitude above Ethereum's~15 transactions per second.\footnote{\url{https://l2beat.com/scaling/summary}}

Prior work has examined the guarantees of correctness and integrity of rollups~\cite{koegl2023attacks,chaliasos2024towards}, including fraud-proof security in optimistic rollups~\cite{lee2025hollow} and soundness issues in validity proofs~\cite{chaliasos2024sok}. In contrast, we investigate \emph{transaction‐fee mechanisms} (TFMs)~\cite{roughgarden2021transaction} that underlie rollup economics. Mis-priced fees can enable denial-of-service (DoS) and finality attacks, preventing transactions from achieving timely L1 finality, disrupting Centralized Exchange~(CEX) withdrawals, bridges, and cross-chain interoperability and settlement, while also inflicting direct financial losses on rollup operators.

A rollup's TFM must simultaneously cover three cost dimensions: (1)~L2 local costs (e.g., execution), (2)~L1 data-availability (DA) (e.g., EIP-4844)~\cite{crapis2024eip}, and (3)~L1 gas for batch settlement and proof verification. 
In this paper, we identify two novel attack classes.  
First, a \emph{DA-saturation attack} floods the sequencer with data-heavy but compute-light transactions, exhausting L1  blob capacity per batch and causing (i)~L2 DoS, (ii)~amplified finality delays, and (iii)~protocol losses.  
Second, a \emph{prover killer attack} that exploits mismatches between Ethereum Virtual Machine gas and proving costs: by crafting (underpriced) high-proof-cost transactions, an adversary delays proof generation and inflicts net losses on rollups.

We evaluated these attacks on major rollups, showing that DA-saturation can be sustained for under~$2$ ETH for~$30$ minutes (and indefinitely for~$0.8$ to $2.7$~ETH per hour in some rollups), and that such \emph{blob-stuffing} via the L2 yields a~$1.45\times$-$2.73\times$ amplification in finality delay over direct L1 blob-stuffing.  
We show that the prover killer attack can delay finality by more than~$8$ hours~($94\times$), while causing~\$$42.26$ in economic damage per block.  
Finally, we propose short- and mid-term mitigations and long-term solutions: a \emph{multi-dimensional} TFM that independently prices L1 DA, L2 execution gas, and proving resources.

\point{Contributions}
\begin{itemize}[leftmargin=*,nosep]
    \item \textbf{Novel DA-Saturation Attacks.} We formalize how a financially-bounded adversary can flood an L2 with data-heavy execution-light transactions to exhaust L1 DA per batch, causing both sustained L2 DoS and amplified finality delays. Notably, this DoS attack affects \emph{all} major rollups in production, significantly hindering the primary goals of a rollup, namely, high throughput and low transaction fees.

  \item \textbf{Prover Killer Attack.} We present the first systematic study of proving-cost mis-pricing in validity proof rollups, identify opcodes and precompiles with the worst cycles-per-gas ratios, and craft high-proof-cost transactions that stall batch proofs and impose net losses on provers. 

  \item \textbf{Experimental Evaluation.} We evaluate both attack vectors on major Ethereum rollups, quantifying attacker expenditure and the resulting impact.

  \item \textbf{Responsible Disclosure and Mitigations.} We reported all vulnerabilities described in this paper to the corresponding rollup teams through their bug-bounty programs, receiving \emph{four} bounties (worth tens of thousands USD). Notably, \emph{all} rollups were susceptible to at least one of the attacks. We collaborated with each team to assess the attacks' impact under their configurations, suggest and document short-term reactive patches and mid-term mitigations, and explored multidimensional fee mechanisms that separately price L1 DA, L2 gas, and proving. All teams applied at least one of the short/mid-term mitigations.
\end{itemize}

\textit{Our work generalizes classical resource exhaustion attacks to the multidimensional setting. We show that mispricing across dimensions introduces novel and practical vulnerabilities absent from prior monolithic blockchains.}

\section{Background}
\label{sec:background}

Scalability is a fundamental challenge that impedes the widespread adoption of blockchains. In highly distributed blockchains such as Ethereum~\cite{wood2014ethereum}, this can lead to a throughput limit, which in turn causes transaction fees to increase during periods of high usage. For example, Ethereum handles around~$15$ transactions per second on average, with demand frequently exceeding this upper bound~\cite{zhou2020solutions}. Numerous techniques have been proposed to scale blockchain performance while preserving security and decentralization~\cite{zhou2020solutions, sguanci2021layer}. Broadly, these either attempt to (1)~improve the L1 (e.g., novel consensus mechanisms) or (2)~develop L2 solutions that move transactions off the L1. In recent years, \emph{rollups} have emerged as a leading approach to scaling~\cite{sguanci2021layer}. 

The core idea underlying rollups is to execute transactions on a L2 but regularly \emph{roll up} batches of transactions and post them on the L1. This allows the L1 to verify and secure a large volume of off-chain activity without requiring any protocol-level modifications. By relocating the bulk of computation off-chain, rollups can achieve significantly higher throughput and lower per-transaction costs while inheriting the security of the L1, since the latter maintains the transaction data and can enforce correctness if the rollup misbehaves~\cite{chaliasos2024towards}.

Rollup designs differ in how they ensure the correctness of off-chain transactions and can be classified into two broad categories: \textit{Optimistic Rollups} and \textit{Validity (ZK) Rollups}. We briefly summarize each category below and then discuss how TFMs are applicable to each. For a broader discussion of rollup security issues, see Koegl et al.~\cite{koegl2023attacks}.

\subsection{Optimistic and Validity (ZK) Rollups} 

\emph{Optimistic Rollups} (ORs) assume that off-chain transaction execution is valid and only engage in dispute resolution if fraud is detected~\cite{kalodner2018arbitrum,bousfield2022arbitrum}. An OR commits a new state root (and the accompanying transaction data) to the L1 without any proof of correctness, therefore operating \emph{optimistically}. To secure this assumption, OR systems have a dispute window during which anyone observing the chain can challenge an invalid rollup state update. If a malicious rollup operator posts an incorrect state, a \emph{fraud proof} can be submitted to demonstrate a specific inconsistency. Upon provision of a successful fraud proof, the rollup contract on L1 reverts the invalid state transition and typically slashes the stake of the party that posted it, thereby economically disincentivizing malicious behavior. Due to this interactive challenge mechanism, ORs provide slow finality: users must wait for the challenge period to be confident that their transactions will not be reverted.

\emph{Validity (ZK) Rollups} (ZKRs)\footnote{ZKRs are more appropriately named validity rollups, since they do not necessarily need to uphold zero-knowledge/privacy.} use cryptographic proofs to assert the correctness of off-chain computations~\cite{WhiteHat_roll_up_token}. Every batch of transactions processed by a ZKR is accompanied by a Succinct Non-interactive Argument of Knowledge (SNARK)~\cite{gennaro2013quadratic} attesting that the new state root was correctly computed from the previous state and the given transactions. Then, a smart contract on the L1 verifies this proof before accepting the state update, providing soundness guarantees~\cite{sguanci2021layer}. 

ZKRs offer better finality than ORs, since transactions are finalized once a proof has been generated, submitted, and verified by the L1. Furthermore, ZKRs leverage compression techniques as only the minimal data required for proof verification has to be posted on-chain. In contrast, ORs require all data to be posted on the L1 to permit the construction of valid fraud proofs. Thus, when using cryptographic proofs, it is possible to post only the outputs of executed transactions, known as \emph{state diffs}~\cite{zksync2023}, to the base layer and verify the proof using only the state diff. However, ZKRs are more challenging to implement securely due to their increased complexity and downstream potential implementation bugs and substantially larger proving costs~\cite{chaliasos2024sok}.

\subsection{Transaction Fee Mechanisms for Rollups}
\label{sec:background-tfms}
Rollups introduce novel challenges for fee mechanism design due to their two-layer nature. A rollup sequencer typically charges users a fee for including their transactions in an L2 batch. This fee must ultimately cover:~(1) the cost of posting the batch data to L1,~(2) any costs of proving and verifying the batch (for ZKRs), and~(3) the sequencer's own operational costs and profit margin. Existing rollup implementations generally use a first-price auction model for transaction ordering. Users submit transactions with a bid (L2 gas price), and the sequencer includes them in descending order of bids until the batch is full. The simplicity of this mechanism can introduce potential issues such as DoS attacks, finality attacks, or economic attacks, as we demonstrate in Sections~\ref{sec:da-saturation} and~\ref{sec:prover-dos}.

Recent research has begun to explore how TFMs can be optimized for rollups~\cite{chaliasos2024pricing,crapis2024eip}. One challenge is that rollups deal with \emph{multi-dimensional resources}: in particular, they consume L1 block space (or \emph{blobs}) for DA, L2 gas (e.g., for execution), and in the case of ZKRs, proof generation. Designing a single fee that accounts for all these resources is non-trivial and subject to ongoing research. Recent work focuses on the unique component of pricing proof computation in ZKRs, which reduces to designing robust \emph{prover markets} for ZKPs~\cite{wang2024proof}. 


\begin{figure}[t]
    \centering
    \includegraphics[width=0.7\linewidth]{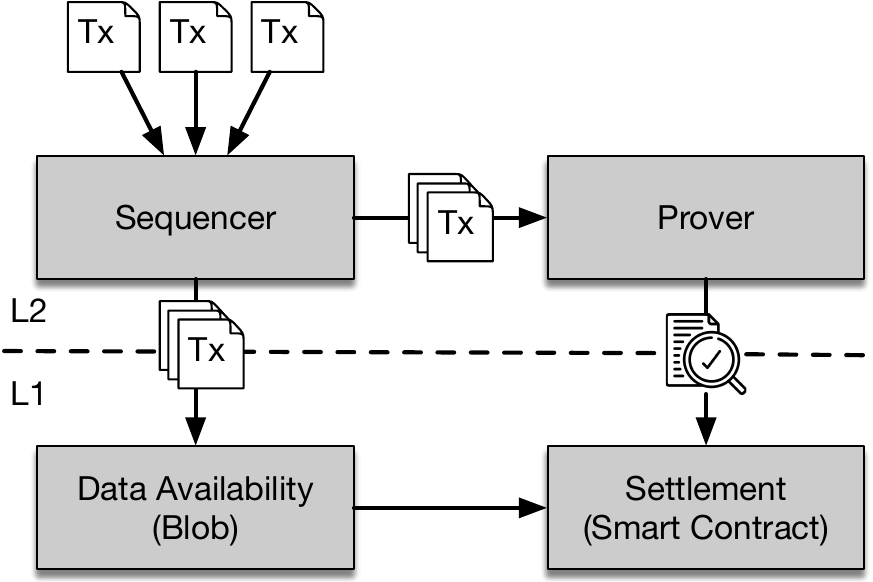}
    \caption{Transaction lifecycle in ZK Rollups.}
    \label{fig:overview}
\end{figure}

\section{Rollup Architecture \& Threat Model}
\label{sec:threatmodel}

\begin{figure}[t]
 \centering
 \footnotesize
 \setlength{\tabcolsep}{2pt}
 \begin{tabular}{@{}ll@{}}
   \toprule
   \textbf{Symbol} & \textbf{Description} \\
   \midrule
   $Tx_{\mathrm{fee}}$ & L2's transaction (tx) total fee \\
   $L2_{\mathrm{fee}}$ & L2's tx fee covering L2's costs \\
   $L1_{\mathrm{fee}}$ & L2's tx fee covering L1's costs \\
   $g$               & L2 gas required by the transaction \\
   
   $\rho_{\mathrm{l2}}$       & Base fee of the current L2 block \\
   
   $\delta$          & Priority fee (tip) offered to the L2 sequencer \\
   
   $\rho_{\mathrm{l1}}$       & Base fee of the current L1 block \\
   
   $C_{\mathrm{tx}}$          & L1 commit cost that needs to be covered by $Tx_{\mathrm{fee}}$ \\
   
   $C_{\mathrm{batch}}$       & L1 commit cost that needs to be covered by a batch \\
   
   $S_{\mathrm{tx}}$          & L1 settlement cost that needs to be covered by $Tx_{\mathrm{fee}}$ \\
   $S_{\mathrm{batch}}$       & L1 settlement cost that needs to be covered by a batch \\
   
   $\rho_{\mathrm{blob}}$     & Blob base fee of the current L1 block \\
   
   $b_{\mathrm{tx}}$          & Bytes required for DA for the transaction \\
   
   \multirow{2}{*}{$\mathit{Scalar}_{\mathrm{blob}}$}   & \multirow{2}{*}{\shortstack[l]{Multiplier for blob fees that adapts the price to include \\ compression benefit and/or potential increases in~$\rho_{\mathrm{blob}}$}} \\
                                      & \\
   
   $bt_{\mathrm{l1}}$         & L1 blocktime \\
   
   $bt_{\mathrm{l2}}$         & L2 blocktime \\
   
   $r$               & Ratio of L1 and L2 blocktimes $bt_{\mathrm{l1}} / bt_{\mathrm{l2}}$ \\
   
   $k$               & Number of blocks \\
   
   $u_{\mathrm{blob}}$        & Rate of increase of blob price between $k_{i}$ and $k_{i+1}$ \\
   \bottomrule
 \end{tabular}
 \caption{List of symbols used in formulas.}
 \label{tab:symbols}
\end{figure}

In this section, we examine the L2 transaction lifecycle, including sequencing, batch sealing, optional proof generation, and L1 settlement. We then formalize a fee mechanism for pricing L2 transactions. Finally, we present our threat model, detailing the adversary's capabilities and security goals.

\subsection{Rollup Execution Path}
Figure~\ref{fig:overview} shows the lifecycle of an L2 transaction in a ZKR. A sequencer first collects transactions and includes them in an L2 batch. It then (i)~posts the batch header and transaction data to L1, so the full L2 state can be reconstructed, and (ii)~forwards the batch to a prover. The prover produces a proof attesting to the correct execution of the batch and submits it to the rollup's L1 contract. Finally, the L1 contract verifies the proof against the previously published data; successful verification finalizes the batch. ZKRs differ in the proving systems they employ and in what they publish to L1; some post full transaction data while others only state diffs~\cite{chaliasos2024analyzing}. ORs differ in that they do not require a validity proof, but instead incur a challenge period to finalize L2 batches.

\subsection{Generic Rollup TFM}
\label{sec:generic-rollup-tfm}
As discussed in Section~\ref{sec:background-tfms}, a rollup’s TFM must (i) charge for the resources a transaction consumes, namely L2 gas, L1 data availability cost, and a share of the fixed L1 gas for committing and verifying batches; and (ii) enable the sequencer to select the most profitable transactions. Users may add a priority tip to incentivize inclusion in the next L2 block. The following \emph{generic} TFM reflects the design of most rollups: A transaction’s total fee is:
\begin{equation}
  Tx_{\mathrm{fee}} \;=\; L2_{\mathrm{fee}} \;+\; L1_{\mathrm{fee}}\,,
    \label{eq:txfee}
\end{equation}
where:
\begin{align}
  L2_{\mathrm{fee}} &= \bigl(\rho_{\mathrm{l2}} + \delta \bigr)\cdot g\,,\\
  L1_{\mathrm{fee}} &= \rho_{\mathrm{blob}} \cdot Scalar_{\mathrm{blob}} \cdot b_{\mathrm{tx}}
                    \;+\; C_{\mathrm{tx}}
                    \;+\; S_{\mathrm{tx}}\,.
\label{eq:l1fee}
\end{align}

Here, $L2_{\mathrm{fee}}$ covers off-chain computation, including gas for execution, storage, bandwidth, and proof generation (for ZKRs), using an EIP-1559-style~\cite{roughgarden2020transaction} base fee $\rho_{\mathrm{l2}}$ plus a priority tip $\delta$, multiplied by the transaction gas $g$. 
Importantly, all L2s operate in competition with other L2s and L1s, striving to minimize their fees.
Note that all existing ZKRs bundle the proving costs into L2 gas.
The on-chain fee $L1_{\mathrm{fee}}$ is comprised of three parts:  
\begin{itemize}
  \item \emph{Data-availability cost}: $\rho_{\mathrm{blob}} \times \mathit{Scalar}_{\mathrm{blob}} \times b_{\mathrm{tx}}$, where $b_{\mathrm{tx}}$ is the size in bytes required for the transaction to be published on the L1's DA, and $\mathit{Scalar}_{\mathrm{blob}}$ accounts for compression benefits and the volatility of blob fees.
  \item \emph{Commit cost} $C_{\mathrm{tx}}$, i.e., the share of the L1 gas to commit this transaction’s data.
  \item \emph{Verification cost} $S_{\mathrm{tx}}$, i.e., the share of the L1 gas to verify the proof for the batch (only for ZKRs).
\end{itemize}
Rollups allocate $C_{\mathrm{tx}}$ and $S_{\mathrm{tx}}$ in different ways (e.g., multiplying the per-transaction cost by a scalar or dividing the batch’s fixed cost across all transactions). Figure~\ref{tab:symbols} lists all symbols we use in the formulas, and Figure~\ref{tab:rollups} (Appendix~\ref{app:tfm-details}) compares how Linea, Polygon zkEVM, zkSync Era, and Scroll implement these components. Notably, zkSync Era posts state diffs rather than full transaction data, over-charges on inclusion, and later refunds any surplus for $C_{\mathrm{tx}}$ and $S_{\mathrm{tx}}$, trading simplicity for robustness. All rollups seal a batch upon reaching a time limit, prover capacity, or data-availability threshold. In all cases, L2 fees remain substantially lower than on L1. 
An overview of L2s, including ORs, appears in Appendix~\ref{app:tfm-details}.

\subsection{Adversarial Model and Security Objectives}
We assume an adversary with all the capabilities available to any L2 user:
they may submit arbitrary transactions, 
use multiple addresses, 
and observe all public L1 and L2 state. 
Adversarial transactions are also indistinguishable from the transactions of honest users. The adversary is constrained by a total budget~$B$ for performing DoS attacks.
We highlight three fundamental security objectives for rollups:
\begin{itemize}
  \item \textbf{Liveness:} every honest transaction is eventually included in an L2 block, i.e., a DoS attack is not viable.
  \item \textbf{Finality:} Once an L2 batch is sealed, it achieves irrevocable confirmation within at most $T$ L1 blocks.
  \item \textbf{Economic Safety:} The L2 must recover all costs of L2 execution/proving and associated L1 fees.
\end{itemize}
\section{Data-Availability Saturation Attack}
\label{sec:da-saturation}

We introduce a novel class of attacks on rollups that exploit the finite DA capacity per batch. An adversary can leverage this constraint to inflict varying degrees of disruption, depending on the DA rules of the rollup, the cost parameters, and the design of the fee mechanism. An attacker with budget $B$ can achieve the following:

\begin{itemize}[leftmargin=*,nosep]
  \item \textbf{Denial of Service.}
    The attacker prevents legitimate traffic from being included by filling a batch's blob quota, occupying all of the DA space, and outbidding honest users.    
  \item \textbf{Amplified Finality Delay.}
    Rather than blob-stuffing the L1 directly, the adversary floods the L2 with DA-heavy transactions, forcing frequent batch seals and prolonging the time until any given batch gets finalized in the L1.
  \item \textbf{Economic Damage.}
    In rollups that amortize the fixed L1 commit cost $C_{\mathrm{tx}}$ over a target number of transactions, the sequencer or protocol treasury must subsidize the shortfall when the attacker’s DA-saturating transactions drive down the effective amortization factor.
\end{itemize}
These attacks leverage two key observations. First, each batch's DA capacity is strictly limited (e.g., by blob count or byte limits), and rollups typically price DA in direct proportion to L1 blob fees, possibly adjusted by a fixed scalar for compression or volatility, without dynamic feedback from L2 congestion. Second, L2 gas fees remain low (the typical target is around~$15$~M gas per block), so submitting hard-to-compress calldata up to the DA limit is economically viable and does not trigger a base fee increase on the L2. 

The severity of the DA saturation attack varies with rollup parameters, such as block times, maximum blobs per batch, L1 commit costs, and prevailing L2 gas prices. 


\point{Rollup Parameters}
Figure~\ref{tab:rollup-comparison} compares the block time, maximum blobs per batch, L1 commit cost (in Gwei), and the median L2 gas price for several prominent rollups. Note that Linea caps each block at a single blob, whereas Optimism throttles DA at peak demand, limiting blocks to~$50$~KB of data and individual transactions to~$10$~KB. Additionally, Arbitrum has a target speed limit of 7M gas per second, enforced through an EIP-1559-like mechanism. Block times range from~$0.25$ to~$3$ seconds, and most rollups allow up to $6$ blobs per batch, but often target at most 3. 

\begin{figure}[t]
  \centering
  \footnotesize
  \setlength{\tabcolsep}{5pt}
  \begin{tabular}{@{}lrrrr@{}}
    \toprule
    \textbf{Rollup}    & \textbf{Block Time (s)} & \textbf{Max Blobs} & \textbf{Commit Cost} & \textbf{L2 Gas} \\
    \midrule
    Scroll             & 3                  & 6                  & 75{,}840             & 0.039                  \\
    Linea              & 2                  & 6 (1)              & 400{,}030            & 0.083                  \\
    ZKsync Era         & 1                  & 1                  & 232{,}524            & 0.045                  \\
    Arbitrum           & 0.25               & 3                  & 168{,}858            & 0.014                  \\
    Optimism           & 2                  & 6 (2)              & 21{,}000             & 0.012                  \\
    Base               & 2                  & 7                  & 21{,}000             & 0.002                  \\
    \bottomrule
  \end{tabular}
  \caption{Comparison of rollup configurations. Max blobs refers to the number of blobs per batch. Numbers in parentheses are DA limits per block: Linea processes one blob per block, and Optimism two. The L2 gas is the mean price between~4/10/2025 and~5/10/2025 in Gwei. Optimism and Arbitrum have threshold and speed limit modes, respectively. Data gathered on May 10, 2025.}
  \label{tab:rollup-comparison}
\end{figure}

\point{Crafting DA‑Heavy, Compute‑Light Transactions}
To maximize DA usage while minimizing the computation, an attacker may issue transactions that execute only the \texttt{STOP} opcode and carry random calldata designed to be compression-resistant. We model the rollup’s compression (e.g., Brotli) by generating uniformly random bytes with a small zero-byte ratio~(3\%), ensuring near-zero compression. Each such transaction consumes minimal L2 compute but fully occupies its DA. Appendix~\ref{app:compression} includes an analysis of compressed DA-heavy transactions and the impact on the cost of our attack if the rollup does not compress the data aggressively. Note that due to the high entropy of the generated calldata, we expect all compression algorithms to perform poorly. Writing a full blob of hard-to-compress calldata requires approximately~$5{,}124{,}950$ gas, so saturating a batch of up to three blobs consumes about~$15.3$~M gas, which is below the typical L2 gas targets.

\point{Conservative Assumptions} 
Throughout this section, whenever we have to make some assumptions (e.g., usage of DA in the L1 beyond the rollup itself), we take worst-case scenarios for the attacker, i.e., minimizing the impact of the attack.
Another critical assumption is that rollups update the blob base fee on every Ethereum block. In practice, most rollups update the blob base fee every $5$ to $64$ L1 blocks, making the attacks much more efficient (see Appendix~\ref{app:delayed-updates}).

\point{Blob Pricing}
Since the introduction of blobs on Ethereum, their base fee has remained extremely low, typically around $1$ wei, with only occasional spikes. If the blob floor price was as high as $0.1$ Gwei, the attack’s effectiveness diminishes dramatically. For instance, under the parameters in Section~\ref{sec:da-dos}, an attacker would be priced out after only $25$ L1 blocks instead of the roughly $150$ blocks required at $1$ wei. For a comprehensive analysis of how L2 ``inscriptions'' temporarily spiked blob prices and impacted L2s in a manner analogous to our attacks, see Gogol et al.~\cite{gogol2024writing}. Note that inscriptions were not deliberate attacks against L2s.  

\point{Note on Arbitrum}
Arbitrum's enforces a hybrid first-come, first-served ordering with additional implementation constraints~\cite{Mamageishvili2023BuyingTL,bousfield2022arbitrum}. Consequently, our DA-saturation attack (especially the DoS attack) is significantly harder to execute on Arbitrum. 
Further, Arbitrum's speed limit also protects against the DoS attack, as the L2 gas price will exponentially grow if more than the target gas is consumed in one second.

\subsection{DA-Saturation Denial of Service}
\label{sec:da-dos}
Liveness is a core property of blockchains: honest transactions must eventually be processed, ideally, without delays. A common liveness attack is denial-of-service~(DoS), which aims to impede the inclusion of transactions. Prior research on TFMs has sought to mitigate DoS by preventing mispricing-induced flooding, e.g., Ethereum’s historical opcode mispricings~\cite{perez2019broken}. However, rollups relying on L1 DA pricing to regulate their bounded DA resource can introduce new attack vectors.

Depending on the configuration of a rollup, a DoS attack based on DA saturation is possible. Intuitively, the idea is that an attacker will consistently consume all the DA space, out-pricing benign users by setting high priority fees, which will result in the L2 not processing benign transactions. The cost of such an attack should be extremely high, making it non-viable for an attacker to perform. However, due to the low gas prices of L2s and the low prices of blob DA, this attack is possible on L2s. 

Importantly, the attack it not a complete DoS. Benign users can still transact by outbidding the attacker on priority fees, which can lead to a higher cost for the attacker to achieve a complete DoS. Still, this bidding war undermines the core value proposition of L2s: low fees and high throughput. Even at L1-level priority fees, the attack remains sustainable, prompting users to migrate back to L1.

Even during short periods, this could lead to attacks with high economic damage, such as arbitrage interference and cross-chain arbitrage~\cite{mazor2024empirical,gogol2024cross}, oracle manipulation~\cite{eskandari2021sok}, and delayed liquidations~\cite{qin2021empirical} from which an attacker can profit significantly. In most cases, the attack eventually becomes non-viable due to the increase in blob price (in L1) due to high blob usage. 

Specifically for Ethereum, the L1 block time ($bt_{\mathrm{l1}}$) is approximately~$12$ seconds, and each block has a target of~$6$ blobs per block and a limit of~$9$ blobs. The maximum increase rate ($u_{\mathrm{blob}}$) between blocks for the blob base fee is ~$\sim1.15$. If, during one L1 block, a rollup consumes more than $6$ blobs, then we can assume that the price will increase and that the attack will eventually become exponentially expensive. As scalar factors (multipliers of blob fee) in L2s can be set by the L2 operators (based on expected compression gains, expected price spikes, etc.), we set the scalars to~$1$. Changing this parameter will not affect our attack. For computing the L2 fees per attack transaction, we will use Eq.~\ref{eq:txfee} from Section~\ref{sec:generic-rollup-tfm}.

We set the priority fee ($\delta$) across the board to~$0.2$ Gwei, and we assume that with~$5.12$M gas, we generate calldata that will fill one blob. We begin with $\rho_{\mathrm{blob}}$ of~$1$ wei, which is the current fee at the time of writing and historically the most common price. If every~$12$ seconds the rollup submits more than $6$ blobs, we assume that in the next L1 block the DA price would have increased by~$1.15$ ($u_{\mathrm{blob}}$). Otherwise, we don't assume that it would increase, although this could happen due to extraneous usage. 

To compute how many blobs the L2 will submit in $bt_{\mathrm{l1}}$, denote the L2 base fee as $\rho_{\mathrm{l2}}$ and the blob fee as $\rho_{\mathrm{blob}}$. The number of blobs submitted per L1 interval is:
\begin{equation}
  T_{\mathrm{blobs}}
  = r \cdot \min\bigl(Block_{\mathrm{limit}},\,Batch_{\mathrm{limit}}\bigr),
  \label{eq:tblobs}
\end{equation}
and the number of transactions is:
\begin{equation}
  T_{\mathrm{tx}}
  = r \cdot \max\bigl(1,\,Tx_{\mathrm{tofill}}\bigr),
  \label{eq:ttx}
\end{equation}
where $r = bt_{\mathrm{l1}}/bt_{\mathrm{l2}}$ and $Tx_{\mathrm{tofill}}$ captures per-transaction DA throttling (e.g., $2$ for Linea and Optimism). The batch rate per interval is:
\begin{equation}
  T_{\mathrm{batches}}
  =
  \begin{cases}
    r & Block_{\mathrm{limit}} = 0\\[0.5ex]
    \displaystyle\frac{q + r\cdot Block_{\mathrm{limit}}}{Batch_{\mathrm{limit}}} & \text{otherwise}
  \end{cases}
  \label{eq:tbatches}
\end{equation}
where $q$ is any carry-over blobs from the previous interval. The total attack cost per L1 interval is:
\begin{align}
  C_{\mathrm{blobs}}
    &= T_{\mathrm{blobs}}\cdot2.08\times10^{6}\cdot\rho_{\mathrm{blob}}
    \label{eq:Cblobs}\\
  C_{\mathrm{calldata}}
    &= T_{\mathrm{blobs}}\cdot2.08\times10^{6}\cdot(\rho_{\mathrm{l2}}+\delta)
    \label{eq:Ccalldata}\\
  C_{\mathrm{txs}}
    &= T_{\mathrm{tx}}\cdot21\times10^{3}\cdot(\rho_{\mathrm{l2}}+\delta)
    \label{eq:Ctxs}\\
  C_{\mathrm{batches}}
    &= T_{\mathrm{batches}}\cdot C_{\mathrm{batch}}
    \label{eq:Cbatches}\\
  \mathrm{Cost}
    &= C_{\mathrm{blobs}}
     + C_{\mathrm{calldata}}
     + C_{\mathrm{txs}}
     + C_{\mathrm{batches}}
    \label{eq:Ctotal}
\end{align}

\begin{figure}[t]
    \centering
    \includegraphics[width=\linewidth]{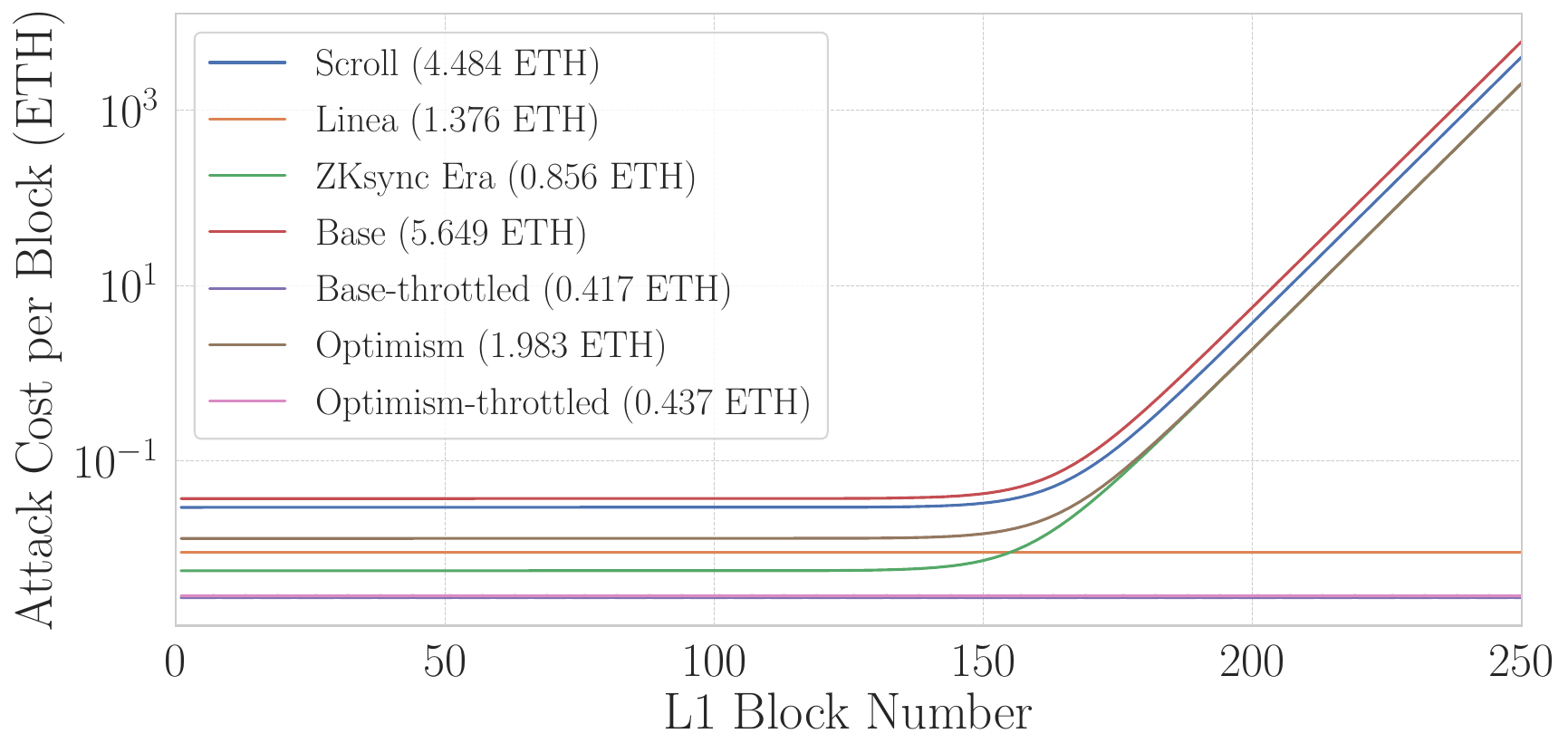}
    \caption{DoS cost per rollup. The cumulative cost for~150 L1 blocks is shown in parantheses.}
    \label{fig:dados}
\end{figure}

Figure~\ref{fig:dados} shows that Linea and Optimism/Base (in throttle mode) are vulnerable to sustained DoS at minimal cost, which means that the cost of the attack does not increase. An attacker can halt the chain by outbidding benign users for a prolonged period (e.g.,~$0.87$~ETH for a DoS attack on the Optimism chain for one hour).
Other rollups can be disrupted for tens of minutes with only a modest cost to the attacker.
Even short disruptions could be catastrophic for chains that secure billions of dollars, both in attack opportunities through cross-chain arbitrage and other economic attacks and reputational damage for the targeted chain.
In Section~\ref{sec:mitigations}, we propose countermeasures to prevent the attack.

\subsection{Amplified Finality Delay Attack}
\label{sec:da-finality}

We extend the attack to delay finality for batches: by filling DA capacity with adversarial batches, benign batches cannot be committed/finalized on L1. We assume that L2s do not absorb any gap between user-paid blob prices and actual fees. Although some rollups initially cover this gap, their fee vaults will eventually deplete, forcing them to opt out and delay batch submissions until L1 blob fees match what users paid.

We analyze how the attack will work in a scenario where an adversarial user performs a blob stuffing attack on the L1, and how the same attack will be amplified if it is performed on the L2.
We show that attacking the L2s for blob stuffing and preventing the finality of benign L2 batches is more effective by at least a factor of~\empirical{$1.45$} when performed through the L2.

\point{Direct L1 Attack}  
The adversary uses a budget $B$ to fill all $b=9$ blobs in each L1 block.  
Let $\rho_{\mathrm{blob},0}$ be the initial blob‐base fee and $u$ the per‐block price multiplier. The cost to saturate all blobs in block $i$ is:

\begin{equation}
c_i = b \cdot \rho_{\mathrm{blob},0} \cdot u^i \quad\text{with}\quad b=9.
\end{equation}
To sustain this for $k_{\mathrm{L1}}$ consecutive blocks, the total budget satisfies:

\begin{equation}
B = \sum_{i=0}^{k_{\mathrm{L1}}-1} c_i
  = 9 \cdot \rho_{\mathrm{blob},0} \cdot \frac{u^{k_{\mathrm{L1}}}-1}{u-1},
\end{equation}
so the attacker can block:

\begin{equation}
k_{\mathrm{L1}}
= \frac{\log\!\Bigl(\frac{(u-1)\,B}{9\,\rho_{\mathrm{blob},0}}+1\Bigr)}{\log u}
\end{equation}
blocks before being priced out.  
Additionally, a cool‐down of $k_{\mathrm{L1}}$ blocks is needed for the base fee to decay back to $\rho_{\mathrm{blob},0}$, for a total disruption of $2k_{\mathrm{L1}}$ blocks.

\begin{figure}[t]
  \centering
  \begin{subfigure}[b]{0.48\textwidth}
    \centering
    \includegraphics[width=\linewidth]{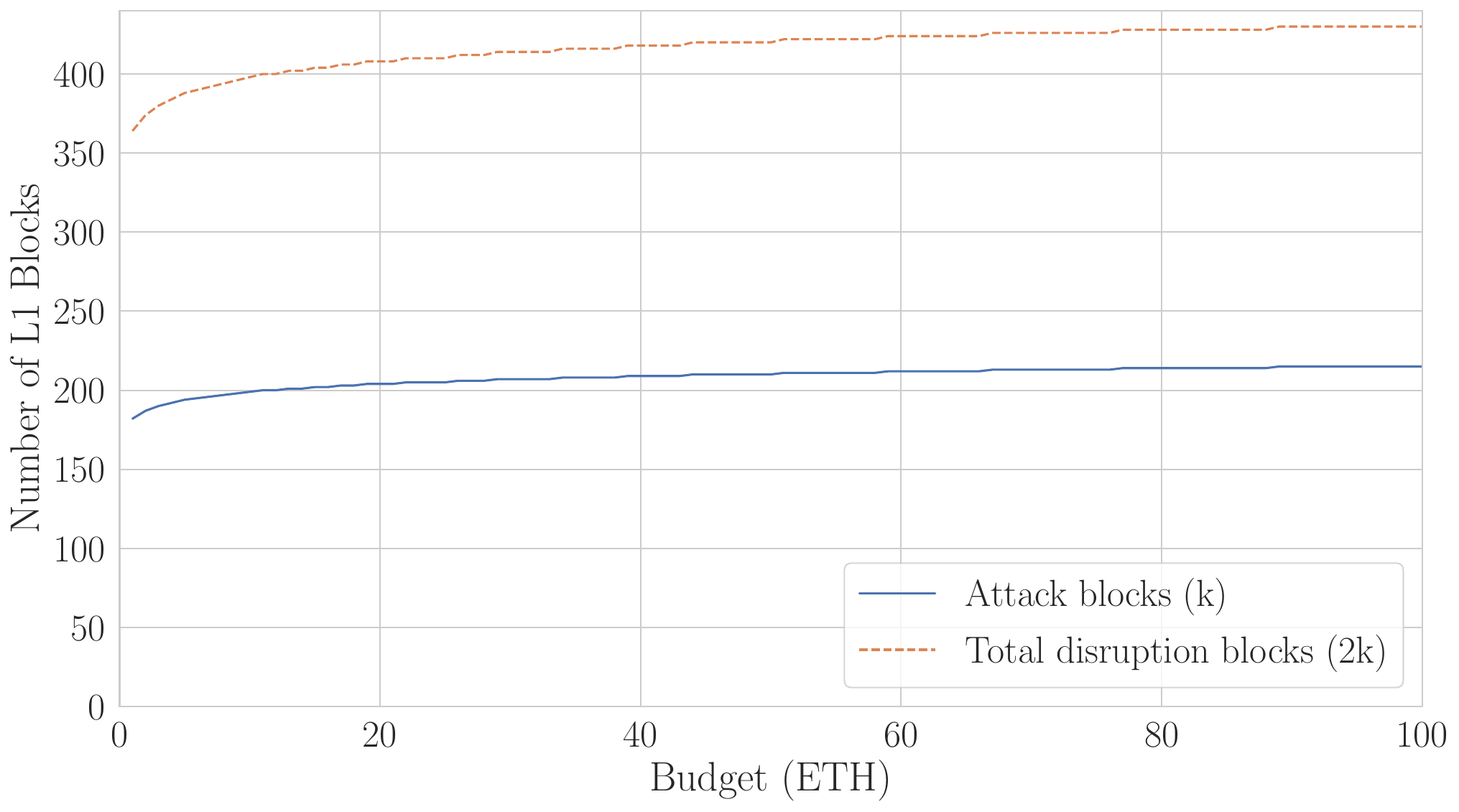}
    \caption{Na\"ive L1 blob stuffing attack.}
    \label{fig:l1da}
  \end{subfigure}
  \\
  \begin{subfigure}[b]{0.48\textwidth}
    \centering
    \includegraphics[width=\linewidth]{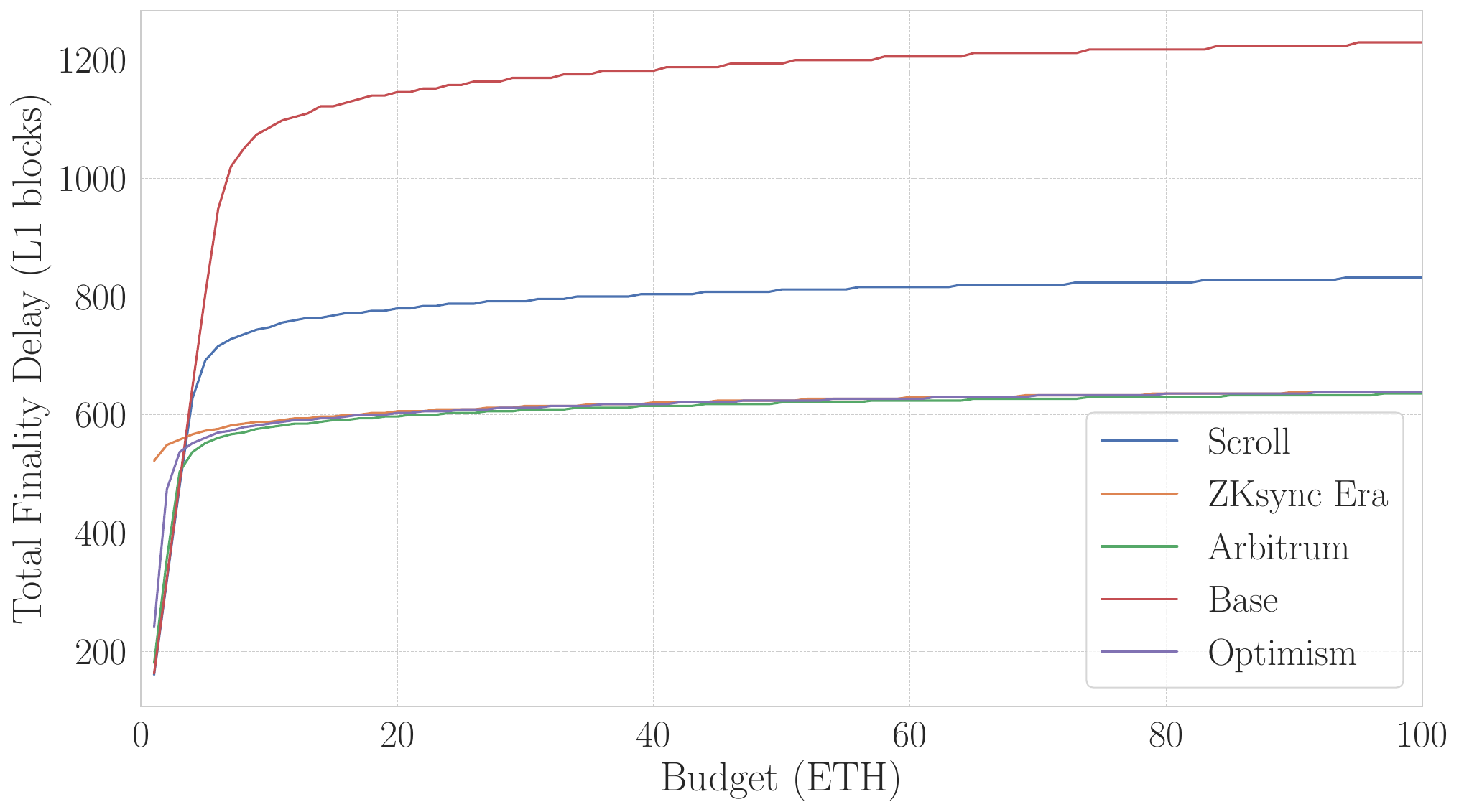}
    \caption{L1 blob stuffing attack via L2.}
    \label{fig:l2da}    
  \end{subfigure}
  \caption{Finality delay attacks through blob stuffing.}
\end{figure}

Figure~\ref{fig:l1da} plots both $k_{\mathrm{L1}}$ and $2k_{\mathrm{L1}}$ for budgets up to $100$ ETH (with $\rho_{\mathrm{blob},0}=1$ wei and $u=1.15$).  
In particular, budgets of~$1$~ETH,~$10$~ETH, and~$100$~ETH yield $2k_{\mathrm{L1}} = 364, \;398,\text{ and }430$, respectively.

\point{Magnified finality delay via rollup attack}
A key insight is that an L2's DA blob-pricing depends almost entirely on the L1 blob base-fee, which is adjusted once per L1 block interval. To keep our analysis simple, we ignore some secondary factors, such as L2-side scalars: in a large-scale DA-saturation attack, these are negligible compared to $\rho_{\mathrm{blob}}$ on the L1.

The attack is most potent when the rollup is forced to commit to multiple batches within a single L1 block interval. In practice, a batch proposer may only be able to post one blob transaction per L1 block, leaving any additional batches queued. We further assume that the L2 operator submits each batch only once the user-paid L2 fee guarantees its inclusion on L1; this can alter the on‐chain ordering of commit transactions, but does not change the fact that batch $i$ cannot finalize until all batches $<i$ are both committed \emph{and} finalized on L1. 

To perform the attack on the L2 level, we compute how many batches the L2 will commit within one L1 block interval ($r$) and the total blobs these batches consume. If this yields at most one batch, the attack is less (or equally) effective as the naive L1 attack.  
We must also account for the L2 calldata fee, along with commit costs and L2 intrinsic cost.

To bound the adversarial advantage and simplify the analysis, we make three worst-case assumptions:
\begin{enumerate}[leftmargin=*,nosep]
  \item Whenever the adversary's batch commit transaction consumes fewer than~$9$ L1 blobs, we assume the remaining blob capacity is filled by others, so the blob base-fee on L1 still increases by the full factor $u=1.15$, making the attack more expensive.
  \item When the L2 is not able to submit a batch transaction because $\rho_{blob}$ is more expensive than what the L2 users paid, no one else submits blobs in that L1 block interval and the blob base-fee decays by 15\%.
  \item L2s update $\rho_{blob}$ every L1 block interval.
\end{enumerate}

Thus, the budget $B$ will be consumed as follows:

\begin{equation}
B = \sum_{i=0}^{k_{\mathrm{L1}}-1} \mathrm{Cost}
  = \sum_{i=0}^{k_{\mathrm{L1}}-1}C_{\mathrm{blobs}, i} + C_{\mathrm{calldata}} + C_{\mathrm{txs}} + C_{\mathrm{batches}}
\nonumber
\end{equation}
where $C_{\mathrm{calldata}} + C_{\mathrm{txs}} + C_{\mathrm{batches}}$, are not affected by L1 price increases and can be grouped into $C_{\mathrm{const}}$:

\begin{equation}
B = \sum_{i=0}^{k_{\mathrm{L1}}-1} C_{\mathrm{blobs},i} + C_{\mathrm{const}}
\end{equation}
Over $k_{\mathrm{L1}}$ blocks, the total budget satisfies:
\begin{align}
B
&= \sum_{i=0}^{k_{\mathrm{L1}}-1} \bigl(b\,\rho_{\mathrm{blob},0}\,u^i + C_{\mathrm{const}}\bigr) 
\nonumber
\\
&= b\,\rho_{\mathrm{blob},0}\;\frac{u^{k_{\mathrm{L1}}}-1}{u-1}
  \;+\;
  C_{\mathrm{const}}\;k_{\mathrm{L1}}.
\end{align}
Solving for $k_{\mathrm{L1}}$, we get:

\begin{equation}
k_{\mathrm{L1}}
= \frac{\log\!\Bigl(1 + (u-1)\,\frac{B - C_{\mathrm{const}}\,k_{\mathrm{L1}}}
                                {b\,\rho_{\mathrm{blob},0}}\Bigr)}{\log u}.
\end{equation}
Since $k_{\mathrm{L1}}$ appears on both sides, we need to solve numerically for $k_{\mathrm{L1}}$, satisfying:

\begin{equation}
b\,\rho_{\mathrm{blob},0}\;\frac{u^{k_{\mathrm{L1}}}-1}{u-1} + C_{\mathrm{const}}\;\cdot k_{\mathrm{L1}}
\;=\; B.
\end{equation}

After the adversary has executed the DA saturation attack for $k_{\mathrm{L1}}$ consecutive L1 blocks, there remain $k_{\mathrm{L1}} \; T_{\mathrm{batches}} - k_{\mathrm{L1}}$ committed but not yet finalized L2 batches.  
Under our simplifying assumptions that each subsequent L1 block that does not contain an L2 transaction and that $\rho_{blob}$ will reduced by~15\%, the rollup can finalize one batch per block, where $\rho_{blob}$ is the same as the one paid by the attacker, then wait for one L1 block for the price to decay and continue.
Consequently, clearing the backlog requires an additional $k_{\mathrm{L1}}$ L1 blocks.  
Including the initial attack phase and the fee-decay \emph{cool-down}, the total finality delay in terms of L1 blocks is:

\begin{equation}
k \;=\; k_{\mathrm{L1}} \;+\; k_{\mathrm{L1}}\cdot\,T_{\mathrm{batches}}.
\end{equation}

Figure~\ref{fig:l2da} shows the result of the attack per rollup. Note that Linea and Optimism in throttling mode are not susceptible to the attack as $T_{\mathrm{batches}} \leq 1$.
The attack successfully achieves a finality delay of~$748$ L1 blocks for Scroll,~$588$ for Era,~$579$ for Arbitrum,~$585$ for Optimism, and~$1{,}086$ for Base, for a cost of~$10$ ETH, achieving at least~$1.45\times$ amplification against the naive attack. Using Base's configurations the amplification becomes~$2.73\times$.

\subsection{Direct Economic Damage Attack}
\label{sec:da-economic}

Beyond DoS and finality delay, an attacker can inflict \emph{direct economic damage} by exploiting how a rollup passes its L1 batch-commit cost \(C_{\mathrm{tx}}\) (see Eq.~\ref{eq:l1fee}) to L2 users.  
Rollups typically allocate \(C_{\mathrm{tx}}\) to L2 by either 
(i)~multiplying each transaction's DA byte-count requirement by a factor to cover \(C_{\mathrm{tx}}\), 
or (ii)~dividing the batch's total DA costs equally among all included transactions.
Both approaches entail some trade-offs depending on the data required to be published.  
For instance, if a rollup publishes state diffs and wishes to charge per-transaction DA bytes, it cannot know each transaction's final size until the batch sealing forces it, leading to \emph{overcharge} up-front and refund later (e.g., zkSync Era).  
Likewise, charging by transaction count requires an initial overcharge with subsequent reimbursement when the number of transactions included becomes known.

If a rollup instead \emph{charges at submission} based on an \emph{expected} or \emph{average} batch size, an adversary can submit DA-heavy, compute-light transactions that incur minimal per-tx fees yet consume full DA capacity.  
The rollup must then subsidize the remainder of \(C_{\mathrm{tx}}\), rapidly draining its vault.  
Scroll was vulnerable to exactly this exploit, suffering an estimated financial loss of~\$$11{,}172$ per hour (Appendix~\ref{app:scroll-poc}).\footnote{This attack was independently discovered by multiple security researchers during the same period.}


\section{Prover Killer Attack on ZK-Rollups}
\label{sec:prover-dos}

\begin{figure*}[t]
  \centering
  \begin{subfigure}[t]{0.35\textwidth}
    \centering
  \footnotesize
    \begin{tabular}{lrrr}
      \toprule
      \textbf{Opcode} & \textbf{Gas} & \textbf{Cycles} & \textbf{C/G} \\
      \midrule
      jumpdest     &     1 &  1039.8 & 1039.79 \\
      difficulty   &     2 &  1192.0 &  596.00 \\
      mulmod       &     8 &  4640.8 &  580.10 \\
      calldatacopy &     3 &  1733.8 &  577.93 \\
      gasprice     &     2 &  1145.8 &  572.88 \\
      address      &     2 &  1144.8 &  572.38 \\
      coinbase     &     2 &  1141.3 &  570.64 \\
      origin       &     2 &  1134.8 &  567.42 \\
      \bottomrule
    \end{tabular}
    \caption{Top 8 highest cycles/gas opcodes.}
    \label{tab:worst-zkvms}
  \end{subfigure}
  \hfill
  \begin{subfigure}[t]{0.28\textwidth}
    \centering
  \footnotesize
    \begin{tabular}{lrrr}
      \toprule
      \textbf{Opcode} & \textbf{Gas} & \textbf{Cycles} & \textbf{C/G} \\
      \midrule
      create     & 32000 &  1679.5 &   0.05 \\
      create2    & 32000 &  1706.7 &   0.05 \\
      sstore     & 21000 &  4927.6 &   0.23 \\
      log4       &  1875 &  2744.2 &   1.46 \\
      log3       &  1500 &  2395.9 &   1.60 \\
      log2       &  1125 &  2291.3 &   2.04 \\
      log1       &   750 &  1914.5 &   2.55 \\
      log0       &   375 &  1604.1 &   4.28 \\
      \bottomrule
    \end{tabular}
    \caption{Top 8 lowest cycles/gas opcodes.}
    \label{tab:best-zkvms}
  \end{subfigure}
  \hfill
  \begin{subfigure}[t]{0.35\textwidth}
    \centering
  \footnotesize
    \begin{tabular}{lrrr}
      \toprule
      \textbf{Precompile} & \textbf{Gas} & \textbf{Cycles} & \textbf{C/G} \\
      \midrule
      modexp                &    200 &  592344.7 & 2961.72 \\
      bn-pair               &  45000 &73896701.8 &1642.15 \\
      bn-add                &    150 &  123223.0 & 821.49 \\
      bn-mul                &   6000 & 4412939.6 & 735.49 \\
      kzg-point             &  50000 & 9490921.2 & 189.82 \\
      identity              &     15 &    1531.0 & 102.06 \\
      sha256                &     60 &    3633.8 &  60.56 \\
      ecrecover             &   3000 &   48942.7 &  16.31 \\
      \bottomrule
    \end{tabular}
    \caption{EVM precompile cycles/gas.}
    \label{tab:precompiles-zkvms}
  \end{subfigure}
  \caption{Prover‐side cycle‐per‐gas benchmarks for ZKVMs.}
  \label{fig:prover-benchmarks}
\end{figure*}

\begin{figure}[t]
  \centering
  \setlength{\tabcolsep}{5pt}
  \footnotesize
  \begin{tabular}{lrrrr}
    \toprule
    & \textbf{Min} & \textbf{Max} & \textbf{Average} & \textbf{Median} \\
    \midrule
    \textbf{Cycles}   & 65.73M & 549.73M & 289.82M & 289.51M \\
    \textbf{Gas}      & 3.48M & 35.85M & 17.70M & 17.85M \\
    \textbf{Cycles/Gas} & 10.41 & 25.98 & 16.61 & 16.48 \\
    \textbf{Proving Time (s)} & 91 & 600 & 319 & 320 \\
    \textbf{Proving Cost, USD} & 0.13 & 0.85 & 0.45 & 0.45 \\
    \bottomrule
  \end{tabular}
  \caption{Statistics for proving blocks from Ethereum's mainnet (with respect to block heights \texttt{22580901}–\texttt{22581000}).}
  \label{tab:normal-blocks}
\end{figure}

This section addresses a unique challenge in ZKRs: the \emph{prover killer} attack. As noted in Section~\ref{sec:background}, each batch in a ZKR requires a validity proof to attest off-chain execution on L1. Yet, proof generation is orders of magnitude more expensive than execution~\cite{ernstberger2024bench}, introducing both higher latency and substantially greater resource consumption. Moreover, certain EVM opcodes incur disproportionately large proving costs compared to their gas cost, creating mispricings under the inherited Ethereum gas metering model. This observation motivates our central question: \emph{Can an attacker exploit mismatches between gas metering and proving overhead to mount latency (finality) and/or economic damage attacks on ZKRs?}

Such an attack would directly undermine ZKR's core advantage of faster finality over ORs by delaying proof production and subsequent L1 settlement. Further, because most EVM-compatible rollups compensate provers solely through accumulated gas fees, without a separate proving fee, a prover may incur net losses if a batch’s transactions demand more proving effort than the collected user fees. 

This concern extends to proposals for \emph{snarkifying} Ethereum. Buterin's suggestion\footnote{\url{https://ethereum-magicians.org/t/long-term-l1-execution-layer-proposal-replace-the-evm-with-risc-v/23617}} to replace the EVM with a RISC-V ZKVM would delegate block proof generation to specialized builders, leaving the rest of the network to perform cheap proof verification. The same vulnerabilities we identify for L2s would then affect the L1. Our aim is to highlight this risk and provide empirical data to guide robust ZKVM and protocol designs.




\subsection{Background: Proving and ZKVMs}

The most popular approach for producing succinct, non-interactive validity proofs of a program's execution is known as a ZKVM~\cite{ben2014succinct}.\footnote{Although the acronym stands for ``zero-knowledge virtual machine'', these proof systems are not necessarily zero-knowledge and can be better described as validity proofs.}
The central idea is to arrange the program execution into a table known as the \emph{execution trace}.
Each row of the execution trace represents one \emph{clock cycle} of a processor, while columns represent registers.
To prove that the program was executed correctly, the prover shows that every transition from one row of the execution trace to the next follows the processor's specification.
Consequently, proving costs increase with the number of cycles being proven~\cite{zkvm-benchmarks2025}.

The above description is an oversimplification of the proof system.
In the same way a hardware CPU is complemented by extra circuits for specialized operations and memory, the execution trace can be accompanied by auxiliary traces for recurring operations (\textit{e.g.}, hash functions) and memory.
Modeling the exact proving costs requires a careful analysis of these auxiliary traces and the cost of proving consistency between trace values.

The general paradigm above can be implemented for any processor architecture.
Directly implementing the EVM~\cite{wood2014ethereum} results in systems known as \emph{ZK-EVMs}~\cite{scroll2023,polygonZkEvm2023,zksync2023}.
A more recent approach consists of implementing a different instruction set and proving the execution of an EVM execution client within the non-EVM processor.
Popular choices of instruction sets include RISC-V~\cite{riscv2016} and WASM~\cite{haas2017bringing} for their existing compiler tooling, or Cairo~\cite{goldberg2021cairo}, and Lita~\cite{thomas2025valida} for their prover-compatibility.
Although this design adds a layer of abstraction, it results in systems that are simpler to maintain, easier to audit, and less prone to subtle bugs, hence their growing adoption as the industry standard.\footnote{All three of zkSync Era, Polygon ZK-EVM, and Scroll have replaced or are planning on replacing their prover with a ZKVM.}

To evaluate the prover killer attack in a representative environment, we focus on a state-of-the-art ZKVM implementation of the EVM --- Succinct's RSP~\cite{succinct2025}.

\subsection{Methodology and Experimental Setup}
Our analysis employs block‐based benchmarking: since L2 batches may aggregate multiple L1‐equivalent blocks, proving cost scales with the number of blocks per batch. Where supported, rollups can use a recursive proof composition to parallelize block proofs~\cite{kothapalli2022nova}, so the finality delay depends on the slowest block plus a recursion overhead.

We use \emph{cycle counts} as a practical proxy for proving complexity.\footnote{\url{https://docs.succinct.xyz/docs/sp1/writing-programs/cycle-tracking}} In a RISC-V ZKVM, each RISC-V operation generally maps to one cycle (though some require multiple), and total cycles directly impact proving time. These metrics are exposed by ZKVM implementations without computing a full proof, making them ideal for identifying proving-expensive operations. However, cycles are only a proxy; outliers may exist, so we do not recommend using them for fee metering. Our methodology proceeds in three phases:

\begin{enumerate}[nosep,leftmargin=*]
  \item \textbf{Cycle Profiling:} We wrote Yul contracts that repeatedly execute each target opcode in different contexts.~\footnote{Contexts are two setups used to get cycles per opcode, where cycles represent internal units of work (e.g., trace rows/constraint evaluations) directly tracking proving time/cost. Contexts were built by profiling 100 consecutive blocks with diverse transactions to estimate per-opcode prover cost. We then wrote micro-benchmarks that repeated each opcode.} 
  By profiling the resulting instruction traces, we obtained stable average cycles per opcode.
  \item \textbf{Mispricing Analysis:} We normalize cycle counts by EVM gas cost to compute a \emph{cycles-per-gas} metric, highlighting opcodes whose proving overhead deviates most from their execution gas.
  \item \textbf{Attack Evaluation:} Guided by the highest cycles-per-gas opcodes and other domain-specific insights, we crafted nine attack scenarios. For each, we measure (a) total cycle overhead, (b) finality delay versus typical blocks, and (c) prover economic loss (proving cost vs.\ collected gas fees).
\end{enumerate}
Note that we focus solely on proof generation. Before producing a proof, the ZKVM must first generate an execution trace; this step is generally faster and can run on smaller, less expensive machines.

All proof generation benchmarks were run on a machine with a single NVIDIA A100~80 GB GPU,~16 vCPUs (6 physical cores),~170 GB of RAM, and~1~TB of disk space, using Succinct's RSP version \texttt{reth-1.3.10}. Dollar cost estimates are based on GCP rates, multiplied by wall-clock time.
Throughout our experiments, we assume an ETH price of~\$$2,500$, prover cost of~\$$5.07$ per hour, and an L2 base fee of~$0.1$ Gwei.

\point{Focus on Reproducibility}
Our setup is intentionally simple so that other researchers can easily reproduce it. In practice, commercial deployments often use more advanced infrastructure (e.g., Succinct's Prover Network with about~$160$ GPUs)\footnote{\url{https://blog.succinct.xyz/sp1-hypercube/}} and potentially better prices. Because our goal is to identify the blocks that are hardest to prove, we expect our relative findings to generalize across different hardware configurations. Further study of parallelization strategies will be necessary to fully understand the attack's impact on more advanced setups.

\begin{figure*}[t]
  \centering
  \scriptsize
  \setlength{\tabcolsep}{2pt}
\resizebox{\textwidth}{!}{%
\begin{tabular}{lrrrrrrr}
    \toprule
    \textbf{Attack} & \textbf{Total Gas (M)} & \textbf{Total Cycles (M)} & \textbf{Proving Time (s)} & \textbf{Proving Cost (USD)} & \textbf{Fees Collected (USD)} & \textbf{Profit/Loss (USD)} & \textbf{Latency Delay (s)} \\
    \midrule
    \texttt{MODEXP}      & 35.49 & 26640.26 & 30,575$^*$ & 43.06$^*$ & 0.80 & -42.26$^*$ & 30255 (+9459.0\%)$^*$ \\
    \texttt{SHA256}      & 35.49 & 2110.15 & 1,501 & 2.11 & 0.80 & -1.32 & 1181 (+369.2\%) \\
    \texttt{BN\_PAIRING}      & 35.51 & 1904.79 & 3,966 & 5.58 & 0.80 & -4.79 & 3646 (+1139.8\%) \\
    \texttt{MCOPY}      & 34.08 & 1061.48 & 808 & 1.14 & 0.77 & -0.37 & 488 (+152.7\%) \\
    \texttt{KECCAK}      & 35.49 & 988.12 & 873 & 1.23 & 0.80 & -0.43 & 553 (+172.9\%) \\
    \texttt{CALLDATACOPY}      & 33.52 & 946.19 & 641 & 0.90 & 0.75 & -0.15 & 322 (+100.5\%) \\
    \texttt{JUMPDEST}      & 34.08 & 864.25 & 602 & 0.85 & 0.77 & -0.08 & 283 (+88.3\%) \\
    \texttt{ECRECOVER}      & 33.52 & 654.72 & 1,729 & 2.43 & 0.75 & -1.68 & 1409 (+440.5\%) \\
    \texttt{BN\_MUL}      & 33.52 & 273.10 & 199 & 0.28 & 0.75 & 0.47 & -121 (-37.7\%) \\
    \bottomrule
  \end{tabular}
}
  \caption{Prover killer attack metrics: total gas, proof-generation time and cost, fees collected, and resulting economic damage in a single block. $^*$: \texttt{MODEXP} values for proving time, fees collected, and latency delay are approximate; RSP crashed after~$10{,}266$ seconds, costing~\$14.46 at that point. The estimates are based on total cycle counts.}
  \label{tab:prover-attacks}
\end{figure*}

\subsection{Measurements}
First, the eight opcodes with the highest cycles‐per‐gas ratios are shown in Figure~\ref{tab:worst-zkvms}, the eight lowest appear in Figure~\ref{tab:best-zkvms}, and all EVM precompiles are summarized in Figure~\ref{tab:precompiles-zkvms}. 

Across all opcodes, the median ratio is~$369.09$ cycles/gas and the mean is~$347.16$ cycles/gas. The most extreme case is \texttt{JUMPDEST} at~$1,039.79$ cycles/gas, followed by~$21$ opcodes in the~$506.84$--$596.00$ range. In contrast, several \texttt{SSTORE}-family operations record ratios below or near~$1$, reflecting their high EVM gas cost and relatively low ZKVM overhead. Among precompiles, heavy cryptographic operations generally incur larger prover overheads, though optimized circuits (e.g., \texttt{SHA256}) and gas‐intensive primitives (e.g., \texttt{kzg-point-evaluation}) exhibit moderate ratios; notably, \texttt{MODEXP} peaks at~$2,961.72$ cycles/gas.

We also collected cycle statistics for a sample of mainnet blocks to establish a baseline for ``normal'' block proof costs (Figure~\ref{tab:normal-blocks}). Although limited in scope, these measurements provide evidence of the real‐world impact of the attacks.

\subsection{Prover‐Expensive Transactions}

We designed several prover killer blocks informed by our cycle‐per‐gas measurements and by known ZKP bottlenecks. Measurement‐guided attacks target the opcodes with the highest cycles/gas ratios; we also include memory‐intensive operations, compute‐heavy cryptographic precompiles, and storage‐access patterns that tend to stress the prover.

To evaluate these, we implemented a contract with a YUL function for each attack. Each function takes an iteration count, letting us run a single or a maximal loop constrained by less than~$36$~M gas cap (typical limit for Ethereum and ZKRs blocks is~$36$~M gas). The main attacks are listed below:

\begin{itemize}[nosep,leftmargin=*]
    \item \texttt{MCOPY}: \texttt{MLOAD} and \texttt{MSTORE} operations to stress test memory heavy workloads.
  \item \texttt{JUMPDEST}: Repeated \texttt{JUMPDEST} instructions to force the prover to validate a large jump‐destination.
  \item \texttt{ECRECOVER}: Frequent ECDSA‐recovery calls (0x01), stressing signature‐verification circuits.
  \item \texttt{EXTCODESIZE} Repeated \texttt{EXTCODESIZE} queries on large contracts to force storage‐access proofs.
  \item \texttt{MODEXP}: Worst‐case parameters to the \texttt{MODEXP} precompile (0x05), maximizing costs.
  \item \texttt{CALLDATACOPY}: Large calldata copies to memory, creating complex memory proofs.
  \item \texttt{BN\_PAIRING}: Intensive BN‐pairing operations.
  \item \texttt{BN\_MUL}: Large‐operand BN‐multiplication.
  \item \texttt{SHA256} and \texttt{KECCAK}: Expensive hash operations.
\end{itemize}

\subsection{Results and Impact}
Figure~\ref{tab:prover-attacks} reports each attack's impact. The \texttt{MODEXP} attack is the most effective, increasing finality latency by~$94\times$ and inflicting a~\$$42.26$ economic loss per block. In future work, we plan to develop a systematic methodology for identifying \emph{optimal} \emph{prover-killer} blocks that maximize either latency delay or prover losses, since these two objectives can diverge; we also plan to adapt our methodology to more ZKVMs.

Our analysis reveals an immediate threat to ZKRs posed by both economic damage and finality delay attacks. Two leading ZKRs have already introduced defenses that trade usability for security:
Era implements custom opcode metering calibrated to proving costs rather than strict EVM equivalence, which substantially mitigates prover‐side attacks but may break contracts that assume EVM semantics. Polygon zkEVM adopts a post‐execution settlement model, charging users based on \emph{actual} proving cost; this approach prevents net losses, but degrades fee predictability at inclusion time and still suffers from potential finality delays. Linea and Scroll, in contrast, rely on standard EVM metering and are fully susceptible to both attack classes. However, their per-block proving limits on certain operations may cap the worst-case finality delays.

\section{Mitigations}
\label{sec:mitigations}
We present defenses against the attacks in Sections~\ref{sec:da-saturation} and~\ref{sec:prover-dos}. We first propose short-term reactive measures, noting the potential to harm honest users. We then outline countermeasures for DA-saturation and prover-killer attacks to preserve liveness, finality, and economic soundness. Finally, we discuss \emph{multi-dimensional} TFMs as a long-term solution, jointly and separately pricing L1 DA, L2 execution, and proving to enable fair, efficient, and attack-resistant fee markets.

\subsection{Reactive Mitigations}
\label{sec:reactive-solutions}

Most rollups today operate through a single centralized sequencer. In this setting, the following reactive mitigations can be quickly deployed to mitigate against both attacks.

\point{Transaction Size Limits}
Enforce a maximum transaction size on DA usage or proving cycles. Any transaction exceeding this threshold is rejected outright. Note that the attack might be even more effective if no additional fees are applied, but only some limits are enforced. For example, Optimism's current throttle cap on bytes per block/transaction makes L2 DA DoS saturation more effective (c.f.\ Figure~\ref{fig:dados}).
\point{Dynamic Fee Scalars}
Apply tiered multipliers on top of the base fee according to a transaction's DA footprint (e.g., $10\times$ for~$1$--$10$\% of a blob,~$100\times$ for~$10$--$100$\%). 
\point{Aggressive Transaction Filtering}
Automatically drop or deprioritize transactions with extremely low compression ratios or those that would strain prover resources, ensuring capacity is reserved for benign traffic.

These measures work in centralized rollups but fail in fully decentralized ones, and risk a perpetual cat-and-mouse game with adaptive adversaries.

\subsection{DA-Saturation Attack Mitigations}
\label{sec:mitigating-da}

The root cause of DA-saturation is that each L2 batch (and hence each L2 block) has fixed DA capacity, yet its price is pegged exclusively to L1 blob fees (plus static scalars).  
Further, because L2 blocks occur every $bt_{\mathrm{L2}}$ seconds, faster than the L1 block time $bt_{\mathrm{L1}}$, local congestion cannot immediately trigger a fee increase, which only happens on the L1.  
This delayed feedback loop enables the DA-saturation attacks.

To close this gap, we propose an `EIP-1559-style' mechanism for L2 DA pricing:
\begin{itemize}[leftmargin=*,nosep]
  \item Maintain an L2 `blob base‐fee' $\rho_{\mathrm{l2}}^{\mathrm{blob}}$ that updates every L2 block based on local DA utilization.
  \item Enforce a floor: $\rho_{\mathrm{l2}}^{\mathrm{blob}}$ must never drop below the current L1 blob fee $\rho_{\mathrm{blob}}$.
  \item When syncing with the L1, if $\rho_{\mathrm{l2}}^{\mathrm{blob}}>\rho_{\mathrm{blob}}$, preserve the higher L2 fee to avoid underpricing.
\end{itemize}
By dynamically adjusting the L2 DA price to reflect real-time utilization, while respecting an L1-fee floor, this mechanism thwarts DA-saturation without relying on centralized throttling or reactive filters.

\subsection{Prover Killer Attack Mitigations}
\label{sec:mitigating-prover}

The root cause of prover-side DoS is that Ethereum's native gas metering does not track the actual cost of generating ZKPs for each opcode in a ZKVM.  
To remedy this, a \emph{prover-aware} metering model is needed:

\begin{itemize}[leftmargin=*,nosep]
    \item \textbf{Empirical Proof System Cost Profiling.}
    For the proof system used, robust profiling (e.g. \cite{ernstberger2024bench, wu2018dizk}) of the costs of its submodules needs to be undertaken.
  \item \textbf{Empirical VM Cost Profiling.}
    Existing empirical approaches to VM profiling (e.g. \cite{perez2019broken, yang2019empirically}), coupled with proof system specific cost analyses, can measure cycle-and-proof costs of each opcode in the target ZKVM to yield per-opcode \emph{proof gas} metrics that reflect real proving time, memory, and energy consumption.\footnote{\url{https://docs.succinct.xyz/docs/sp1/optimizing-programs/prover-gas}}
  \item \textbf{Calibration and Validation.}
    Validation of the profiling results against end-to-end proof time and prover resource consumption on representative hardware, ensuring the above correlate with operational costs.
\end{itemize}

Although the above constitute an important first step to understanding the cost profiles of opcode execution, accurate per-opcode proving costs are fundamentally argument-dependent and thus require \textit{worst-case} analyses of the potential costs incurred. Moreover, it is not clear that VM cycles (or any VM-specific datapoints) are the `right' metric to model proving costs. Indeed, costs for proof generation are fundamentally dependent on proof system metrics, which vary widely by the specific type of proof system used and, in most cases, are open research questions themselves. Given the above, two integration strategies are possible:

\begin{enumerate}[leftmargin=*,nosep]
  \item \textbf{Unified Gas Schedule.}
    Incorporate proving gas into the L2 gas, as in Era's updated metering. This preserves a single gas abstraction but sacrifices EVM-equivalence, since opcode costs will diverge from Ethereum's.

  \item \textbf{Dedicated Proving Fee.}
    Introduce a separate $\mathit{L2ProvingFee}$ dimension, akin to the $\mathit{L1Fee}$ term in Eq.~\ref{eq:l1fee}, that users pay per transaction to cover proof generation. This approach maintains EVM-compatibility while cleanly isolating proving costs in a multidimensional TFM, explored in Section~\ref{sec:multidimensional-tfm}.
\end{enumerate}


\subsection{Multi-Dimensional TFMs for Rollups}
\label{sec:multidimensional-tfm}

Although attack vectors through maliciously crafted transactions have been extensively studied in previous work~\cite{chen2017adaptive, perez2019broken}, with mechanisms such as EIP-1559 being developed to mitigate such concerns~\cite{roughgarden2020transaction}, the attack described in this work is applicable mainly due to a failure in TFM design over multiple underlying resource parameters. Specifically, traditional block stuffing attacks in Ethereum have targeted execution gas usage (which has historically been the only underlying resource of concern). In our case, the existence of separate resources provides an additional degree of freedom for a malicious attacker to bypass the safeguards inherent in EIP-1559. To resolve this attack in a generic manner, we need to consider how all underlying resources contribute to pricing and incorporate them into any proposed solution.


This problem is not new and has been modeled in prior work by using multidimensional fees~\cite{diamandis2022dynamic}. In this context, the resource footprint of a potential block can be priced by looking at an indicator function $x \in \{0,1\}^n$ over $n$ potential transactions, where $x_j$ refers to whether the $j$-th transaction is included in the block. We can then construct a matrix $A \in \mathbb{R}^{m \times n}$ over $m$ scarce resource parameters (which in our case are $4$, gas, DA, proving, fixed costs) whose $j$-th column $A_j \in \mathbb{R}^m$ represents the resource costs of the $j$-th transaction, with $j \in [n]$. This yields a resource cost of $Ax$ and can be used to provide precise metrics based on which pricing mechanisms can be designed. 

This formalism is useful, as it allows the inclusion of transactions in a block based on a predefined utilization of \textit{ targets} $b^{*},$ thus aiming to minimize $Ax - b^{*}$ over $x$ when building blocks for submission. This constraint can be furthered by imposing an upper bound per resource, namely requiring that $Ax \leq b$ for some \textit{resource limit} $b$. Note that in the univariate case, the resource is gas while the target utilization parameter is the target gas usage, and the resource limit is the maximal gas usage permitted per block.

The mitigations in the previous subsection can be reformulated as imposing specific network fees \textit{per resource} while retaining network performance. This is equivalent to designing a process to choose the network fee pricing vector $p \in \mathbb{R}^m$ so that $p^TAx$, or the total price of the block, is adequate to prevent attacks while minimizing transaction costs. In the initial EIP-1559 specification, an adaptive update to these parameters (but in the one-dimensional setting) is proposed, written below using the above language:
\[
p_{t + 1} = p_t \cdot \left( 1 + \frac{1}{8} \cdot \frac{(Ax)_t - b^{*}}{b^{*}}\right)
\] 
where $p_t$ is the network base fee and $(Ax)_t$ the total gas used in block $t$. In this paradigm, base pricing of gas (or any multidimensional network fee $p$) is proportionately updated at every block based on its deviation from a pre-specified target. Over time, any attack vector would be resolved by an adaptive increase in the network price, as the system is sensitive to overshooting the target and can also lower fees when the network is underutilized. This strikes a balance between performance and attack mitigation, although it has only been leveraged in the univariate setting.

Thus, a potential solution is multidimensional modeling, which ensures that we can iteratively arrive at adequate base fees for all resources while still preventing attacks. 
The key requirement is that network fees are (or adaptively become) high enough to prevent attacks while, ensuring that they do not make the network too costly to use for honest parties.

We conclude by noting that simply applying the aforementioned EIP-1559 rule independently element-wise per resource \cite{Buterin2025MGP} does not fully resolve the specified concerns. One problem with this is that the complex interdependence between the underlying resource parameters means that it would be difficult to arrive at performant solutions that would adaptively discover optimal prices without additional context on the dynamics between the given variables. 


\section{Discussion}
\label{sec:discussion}


\point{Asymmetric Resource Exploits in Multi-Dimensional Systems}
This paper introduces two novel classes of asymmetric resource exhaustion attacks that exploit emerging design patterns in rollup-based blockchains (and beyond, e.g., L1s with ZK). First, our \emph{DA-saturation attack} targets the economic design of L2 systems that price data availability (DA) based solely on the parent L1's pricing mechanism. This architecture introduces a unique vulnerability: when DA costs are externally priced but locally consumed, malicious actors can saturate the L2's DA capacity at low cost, bypassing standard fee-based congestion controls. This decoupling of pricing and consumption is intrinsic to modular rollups and opens a new attack vector absent from traditional monolithic chains. Second, our \emph{prover-killer attacks} extend classical gas-resource mismatch attacks (e.g., EIP-150) into the multi-dimensional context of ZK rollups. Unlike prior work that focuses on CPU/memory/storage usage, we profile opcode-specific \emph{proving costs} -- capturing algebraic complexity rather than execution time -- and show how cheap transactions can induce high prover effort. This divergence between L2 gas and prover cost is novel in practice and methodology: prior replay-based exploit generators~\cite{perez2019broken} cannot detect prover mispricings without integrating a proving-cost oracle. Our findings highlight the need for multi-resource-aware fee models and lay the foundational ground for future exploit detection in systems with heterogeneous cost dimensions.

\point{Attack Applicability to Alternative DA Layers} 
In case of Celestia~\cite{celestia2023}, L2s could experience an even worse attack where cross-rollup congestion occurs and rollups find their transactions delayed or squeezed out because an attacker has consumed the block space. Data availability sampling~(DAS) \cite{al2018fraud}, one of Celestia's key innovations, does not alleviate this form of attack. DAS helps light clients verify that published data is available, but does not reduce the burden on the network to carry large volumes of data. Every full node and validator would still need to process and store the attacker's blobs, at least temporarily. Thus, while Celestia's architecture is designed for high throughput, it remains susceptible to economic spam attacks if an attacker is willing to burn funds on fees. The qualitative risk is moderate: Celestia's higher throughput (relative to Ethereum) means an adversary can do more damage per block, but conversely, it must implement correspondingly robust fee dynamics or throttling.
In general, there do not appear to be any reasons why DA layers with higher throughput would make the attack less feasible. In contrast, it might be cheaper if the underlying rollups have lower costs for posting data in the L2. One critical observation is that, depending on the limit of DA space per block, if the attacker needs to produce more hard-to-compress calldata (in terms of gas) than the target gas in the L2, then the gas price of the L2 would go up making the attack more expensive. Notably, alternative DA solutions typically also increase the gas limit.

\point{Broader Ecosystem Impact of DA Attacks} 
A successful DA-layer DoS attack has impact beyond just the immediate target rollup, it can ripple across the entire rollup ecosystem and applications sharing that DA layer. This is a key concern in a modular blockchain paradigm: many rollups may outsource their data availability to a single network (be it Ethereum, Celestia, etc.), forming a multi-tenant environment. Under such shared conditions, an attack on one tenant effectively becomes an attack on all. For example, if an attacker manages to consistently fill Ethereum’s blobs, every rollup that uses Ethereum for data availability will be impacted. Their state update transactions might be delayed or they might have to pay much higher fees to outbid the attacker for inclusion. In a worst-case scenario, if the DA layer's capacity is totally saturated, some rollups could even halt progression (since they cannot publish new state data securely) until the congestion is resolved. 

Shared DA layers thus introduce a common problem: the security of each rollup is entwined with the others in terms of data throughput. An attacker might target the most vulnerable rollup on a DA chain, knowing that the resulting congestion will impact all the others as collateral damage, providing a potentially more efficient way to cause widespread disruption than attacking each rollup in isolation. Moreover, if one rollup has weaker fee defenses (say it undercharges for data, effectively subsidizing attacks), an attacker can exploit that rollup as a \emph{Trojan horse} to spam the DA layer cheaply while the damage propagates to others. 

This underscores the importance of coordinated defenses. All rollups on a shared DA service should ideally adopt strong anti-DoS measures, and the DA layer itself should enforce fair pricing. In the broader ecosystem, a severe DA-oriented attack can also undermine confidence in the scaling model. If, for example, Celestia was regularly clogged by attackers, rollup developers might hesitate to rely on them, or users might question the reliability of rollups vs. using an L1 directly. In summary, DA-oriented DoS attacks have a magnifier effect in multi-rollup ecosystems. They threaten not just a single chain, but potentially an entire class of L2s. 

\point{Trade-offs between Usability and Security}
A recurring theme in this work is the fundamental trade-off between usability and security.  
For example, maintaining strict EVM-equivalence via the standard gas schedule simplifies deployment but leaves rollups vulnerable to prover killer attacks, whereas custom gas metering mitigates that risk at the cost of compatibility.  
Similarly, overcharging and refunding ensure accurate cost recovery but could frustrate users, whereas upfront approximations improve UX but open the door to economic attacks.  
Rollup designers must navigate these decisions to remain competitive against alternative L1s, but short-term gains can yield long-lasting reputational harm if attackers exploit the weaknesses identified in this work.  
We believe that a thoughtfully designed, multidimensional TFM, separately pricing DA, execution, and proving, offers a promising path to reconcile security with usability.  

\point{Implications for Snarkified L1 Designs}
The findings of Section~\ref{sec:prover-dos} highlight a critical challenge for any \emph{snarkyfied} L1 design, such as Ethereum's proposed RISC-V ZKVM roadmap. Unlike Ethereum's gas metering, proving costs vary dramatically across different prover implementations and proof systems. With client and prover diversity, accurately pricing proof generation, particularly in a multidimensional fee mechanism, becomes highly complex.

\section{Related Work}
\label{sec:related}

\point{Transaction Fee Mechanism Design}
TFMs have been extensively studied through the lens of economic mechanism design. Early blockchains such as Bitcoin and pre-1559 Ethereum~\cite{leonardos2021dynamical} relied on simple first-price auctions for transaction inclusion, which can lead to inefficient fee allocation and unpredictable costs. Recent research formalizes TFMs and proposes improved auction formats. Roughgarden~\cite{roughgarden2021transaction} introduced incentive-compatibility notions (e.g., miner-mechanism incentive compatibility) tailored to blockchain settings and proved that Ethereum’s EIP-1559 fee market satisfies strong truthfulness and collusion-resistance properties under most conditions. Chung and Shi~\cite{chung2023foundations} developed a theoretical foundation for TFMs, characterizing conditions for user and miner-proofness. Beyond Ethereum's base fee model, recent research has explored multi-dimensional and dynamic fee markets. Diamandis et al.~\cite{diamandis2022dynamic} propose pricing multiple resource dimensions (such as computation vs. data availability) with a mechanism generalizing EIP-1559 to multi-resource environments. In addition, optimal control and game-theoretic approaches have been applied to the dynamics of fees: Crapis et al.~\cite{crapis2024optimal} derive optimal time-varying fee policies to efficiently allocate block space under fluctuating demand, showing that dynamic fees can substantially improve throughput and fairness over static gas limits. These efforts collectively aim to balance user cost minimization with validator revenue and network security. More recently, Wang et al.~\cite{wang2025framework} present a unified framework that jointly models L1 posting strategies and L2 fee dynamics, proving threshold-based policies for batch submission and optimal L2 fee adjustment rules under volatile L1 gas prices and L2 queue congestion. These efforts collectively aim to balance user cost minimization with validator/operator revenue and network security.

Our work complements ongoing research on multidimensional transaction fee mechanisms by demonstrating practical attacks on existing rollups that use naive TFMs. Although multidimensional TFMs offer a promising direction for addressing the complexities of rollup fee design, they have yet to be deployed in practice, highlighting the urgency of understanding current vulnerabilities and designing TFMs that can address them.

\point{Benchmarking and Analyzing ZK-Rollups}
Zero-knowledge proof-based systems, particularly ZKRs, have received significant attention as a scalability solution for legacy blockchains such as Ethereum. Consequently, recent work has focused on benchmarking their performance and understanding their technical trade-offs. Chaliasos et al.~\cite{chaliasos2024analyzing} present a comprehensive analysis of ZKRs, both theoretical and empirical, comparing different rollup designs and quantitatively evaluating their on-chain costs and prover time. Followup theoretical work further presents the main challenges and properties required for rollup TFMs~\cite{chaliasos2024pricing}. Our work builds on that research and presents the first practical attacks on rollup TFMs. 

\point{Denial-of-Service Attacks on Permissionless Blockchains}
Blockchain networks have faced numerous DoS attacks targeting their consensus, networking, and execution layers~\cite{raikwar2021attacks,chaganti2022comprehensive}. A prominent early example is the~2016 Ethereum Shanghai attacks, where an attacker exploited under-priced opcodes to flood the network with computationally heavy transactions, degrading performance and forcing emergency protocol updates. Chen et al.~\cite{chen2017adaptive} analyzed these incidents and proposed an \emph{adaptive gas cost} mechanism that dynamically adjusts opcode pricing to mitigate such DoS vectors. Similarly, Perez and Livshits~\cite{perez2019broken} developed a methodology to detect mispricing in Ethereum's gas metering mechanism and identify potential exploits. 
Their mechanisms target ordering and network-latency markets rather than proving-cost heterogeneity, so it is not directly applicable as-is for finding prover-killers. Yet, it could be adapted by adding a proving-cost dimension (e.g., prioritization rules that internalize cycles-per-gas), which we leave as promising future work.
He et al~\cite{he2024nurgle} shows how to exacerbate \emph{execution-layer state-storage} costs via MPT manipulation, within a one-dimensional gas model that prices CPU/memory/storage work during \emph{execution}.
However, their approach does not address prover-cost asymmetries.
A potential adaptation is to map MPT-heavy access patterns to \emph{prover-cost signals} in zkEVM/zkVM circuits (e.g., per-node/Keccak constraint counts or witness size) and incorporate these into prover-aware fees or prioritization.
Beyond protocol mispricing, researchers have identified incentive-layer DoS attacks that do not rely on volume spam but on strategic manipulation of miner behavior. Mirkin et al.~\cite{mirkin2020bdos} describe the \textit{BDoS} attack, an incentive-driven blockchain DoS in which a malicious miner withholds its blocks to shrink honest miners' rewards and discourage them from mining. In the context of Ethereum’s recent fee and proposer/builder separation model, Yaish et al.~\cite{yaish2024speculative} demonstrate \emph{speculative} DoS attacks that exploit transaction execution semantics and mempool logic. They craft sequences of dependent transactions that appear valid (and attractive to block builders) but ultimately revert or become invalid, thereby evicting honest transactions from mempools or wasting block builder resources. These works underscore adversaries continually adapting DoS strategies as blockchains evolve (e.g., new fee mechanisms). Similarly, rollups introduce new dimensions that attackers can exploit to perform DoS attacks, as demonstrated by the attacks presented in this paper. 

\point{Prover Killers}
Previous work in EthProofs~\cite{ethproofs_website}, a project that benchmarks ZKVM performance for Ethereum, has identified \emph{organic} prover-killer blocks~\cite{ethresearh_prover_killer_blocks}, where benign transaction blocks naturally incur high proving times. In contrast, we construct \emph{adversarial} prover-killer attacks, deliberately crafting blocks to maximize proof complexity. The fact that organic prover killers already exist implies that naive filtering could reject legitimate blocks, underscoring the need for more fundamental mitigations.
In addition, the Polygon team has developed an EVM toolchain to stress test their prover~\cite{agglayer_evm_stress}. We drew inspiration from their YUL contracts and adapted them to stress-test a ZKVM-based implementation. Our goal is to complement existing benchmarking and stress-testing efforts with additional insights.

\section{Conclusions}
\label{sec:conclusions}
We investigated \emph{two new attack classes} that exploit mispriced resources in today's rollups.  
\emph{DA-saturation attacks} flood a rollup with data-heavy, compute-trivial transactions; an adversary can:  
\emph{i)} halt user transactions for \,$\approx 30$ minutes on most rollups with around $2$ ETH,  
\emph{ii)} keep three rollups in continuous DoS for about $0.8$ to $2.7$ ETH/h, and  
\emph{iii)} stretch L1 finality by $1.45\times$–$2.73\times$ relative to naive L1 blob-stuffing.  
\emph{Prover-killer attacks} craft high-proving-cycle blocks that stall ZK proof generation; a single $35$ M-gas block can raise finality latency by $94\times$ and burn \$$42.26$ more in proving costs than it collects in fees.
We disclosed these issues to all affected teams and proposed several mitigations.  
With \$$55$ B already secured by rollups, adopting resource-aware pricing and standardized prover-gas models is now critical to ensure liveness, fast finality, and sustainable economics across the entire rollup stack.

\section*{Acknowledgments}
We want to thank all the rollup teams for engaging with us to resolve the presented vulnerabilities, and especially Matter Labs (zkSync Era), Consensys (Linea), Optimism, and Tools for Humanity (WorldChain) for their generous bounties.  
Furthermore, we would like to express our gratitude to 
Luca Donnoh (L2Beat),
Emiliano Bonassi (Conduit),
Sebastian Stammler, George Knee, Kelvin Fichter (Optimism),
Jan Gorzny, Martin Derka, Valerian Callens, Pavel Shabarkin (Zircuit),
P\'eter Garamv\"olgyi, Alejandro Ranchal-Pedrosa (Scroll),
Steven Smith (Tools for Humanity),
Daniel Gretzke, Joshua Gutow (Uniswap),
Ben Berger, Tsahi Zidenberg, Akaki Mamageishvili (Offchain Labs),
Julien Marchand, Jones Ho, Valentin W\"ustholz (Concensys),
Ignacio Hagopian, David Theodore (Ethereum Foundation),
and Stas Breadless (Matter Labs) 
for their insightful discussion and valuable feedback on the attacks and defenses presented in this work.
Finally, we thank the Latest in DeFi Research (TLDR) program, funded by the Uniswap Foundation, for its support of this work.

\bibliographystyle{plain}
\bibliography{main}{}

\appendix

\begin{figure*}[t]
\centering
\footnotesize
\setlength{\tabcolsep}{5pt}
\renewcommand{\arraystretch}{1.1}
\begin{tabular}{lllll}
\toprule
\textbf{Item} & \textbf{Linea} & \textbf{Polygon zkEVM} & \textbf{zkSync Era} & \textbf{Scroll} \\
\midrule
\textbf{L1 Commit Transaction}              & Bounded to $b_{\mathrm{tx}}$ & Sustainability Margin       & Overhead share                            & Bounded to $b_{tx}$            \\
\textbf{L1 Verify Transaction}              & N/A                          & Sustainability Margin       & Overhead share                            & Covered by $\rho_{l2} \cdot g$ \\
\textbf{DA Data}                   & Transaction Data                      & Transaction Data                     & State Diff                                & Transaction Data                        \\
\textbf{L1 DA fee}                     & $\rho_{\mathrm{blob}} \cdot b_{\mathrm{tx}}$   & $\rho_{\mathrm{blob}} \cdot b_{\mathrm{tx}}$  & $\rho_{\mathrm{blob}} \cdot b_{\mathrm{tx}} \cdot \mathsf{overhead}$ & $\rho_{\mathrm{blob}} \cdot b_{\mathrm{tx}}$     \\
\textbf{L2 Execution Gas}          & Constant                     & Proportional to $\rho_{\mathrm{l1}}$ & Similar to EIP-1559                    & Similar to EIP-1559                  \\
\textbf{Gas Metering}              & L1                  & L1                  & Custom                                    & L1                    \\
\textbf{Proving Transaction}                & Constant                     & Scalar Multiplied by $g$    & Covered by $g$                            & Covered by $g$                 \\
\textbf{Fee Lock}             & Transaction Submission                & Repriced in Sequencer       & Batch Seal                                & Transaction Submission                  \\
\textbf{Refund}                    & N/A                          & Partial                     & Full                                      & N/A                            \\
\textbf{Priority}                  & $\delta$                     & $\delta$                    & $\delta$                                  & $\delta$                       \\
\textbf{Affected by L1 Volatility} & L2                           & L2 and User                 & User                                      & L2                             \\
\bottomrule
\end{tabular}
\caption{Comparison of TFMs across ZKRs. For brevity, we omit the  $\mathit{Scalar}_{\mathrm{blob}}$ factor in the L1 DA fee.}
\label{tab:rollups}
\end{figure*}

\section{Transaction Fee Mechanisms of Major Rollups}
\label{app:tfm-details}
Figure~\ref{tab:rollups} outlines divergences in the TFMs of ZKRs. Specifically, each rollup is characterized based on:

\begin{itemize}
\item \textbf{L1 Commit Transaction}: How the rollup recoups the L1 gas spent to post commit a batch. Implementations typically (i) add a fixed overhead per transaction or (ii) charge the cost across transactions in proportion to their DA-byte requirement.

\item \textbf{L1 Verify Transaction}: This criterion addresses how each rollup accounts for verification costs incurred on L1. Notably, Scroll integrates this cost into L2 gas fees, while Linea currently subsidizes verification costs entirely.

\item \textbf{DA Data}: Specifies the type of data each L2 publishes to L1. Notably, Era differs by publishing state diffs, which can only be accurately computed at batch sealing (though an overapproximation is possible during transaction submission). Other rollups typically publish complete transaction data required for successful chain reconstruction.

\item \textbf{L1 DA fee}: Represents the mechanism used to charge transactions for L1 data availability. All rollups employ a pricing structure based on multiplying transaction DA requirements by the L1 blob base fee, often combined with additional scaling factors.

\item \textbf{L2 Execution Gas}: Defines how each rollup sets its L2 gas base fee. Linea employs an EIP-1559-like mechanism designed to stabilize the base fee, whereas Polygon zkEVM dynamically aligns fees proportionally to L1 costs. Both Era and Scroll utilize standard EIP-1559-like mechanisms.

\item \textbf{Gas Metering}: Each rollup slightly modifies Ethereum's gas metering to accommodate differences in opcodes and precompiles. Era, however, significantly redefines gas metering to directly account for proving complexity and associated costs.

\item \textbf{Proving Transaction}: Captures how each rollup charges for transaction proving. Linea sets a constant fee, Polygon zkEVM applies a scalar multiplier to gas usage, while Era and Scroll incorporate proving costs into their standard gas fees.

\item \textbf{Fee Lock}: Reflects when transaction costs become fixed or known to users. Rollups either lock fees at transaction submission (providing predictable costs) or initially overcharge and subsequently issue refunds based on actual incurred costs. This highlights a trade-off between user experience, security, and implementation complexity.

\item \textbf{Refund}: Related to fee locking, this property indicates rollups that intentionally overcharge at transaction submission and later issue partial or full refunds based on actual costs.

\item \textbf{Priority}: All rollups enable users to specify a priority fee, incentivizing sequencers to include transactions promptly in upcoming blocks.

\item \textbf{Affected by L1 Volatility}: Identifies who bears the short-term risk and financial impact of L1 price volatility, which can either be the rollup operator (sequencer) or end users.
\end{itemize}

Below, we provide some further details for prominent ZKRs (Linea, Polygon zkEVM, zkSync Era, Scroll) and ORs (Optimism, Arbitrum). For brevity, Optimism and Arbitrum are excluded from Figure~\ref{tab:rollups}. As ORs, their fee models are simpler and quite similar. Both commit transaction costs proportional to byte usage, incur no explicit L1 verification costs, publish transaction data directly, base DA fees on L1 prices, utilize EIP-1559-like mechanisms for L2 base gas pricing, employ gas metering standards similar to Ethereum's, have no proving costs, lock fees at transaction submission, issue no refunds, permit priority fees, and shift short-term L1 price volatility risks to the sequencer. A more thorough analysis of each rollup's mechanism follows in subsequent sections.

\subsection{Linea}
Linea charges users via an L2 execution fee (a constant base plus a user-specified priority fee) and an L1 data fee computed dynamically on a per-byte basis based on recent L1 gas conditions. Fees are calculated and locked at transaction inclusion time, meaning that any subsequent volatility in L1 gas prices is absorbed by the sequencer. Linea does not implement refunds after transaction inclusion, and user priority is directly determined by the priority fee set by users. Regarding proving, Linea charges a constant fee and subsidizes verification costs. This model provides simplicity and predictability but has volatility risks~\cite{lineaFees}.

\subsection{Polygon zkEVM}
Polygon zkEVM's fee structure integrates L2 execution costs pegged at approximately $4$\% of L1 execution gas cost (adjusted dynamically) and data posting costs to L1. A distinct feature, the Effective Gas Price (EGP) mechanism, dynamically adjusts fees to ensure transactions with significant calldata requirements pay proportionally more. Partial refunds are available if fee estimations significantly exceed actual costs. Further, Polygon calculates exact proving cost during batch sealing. Notably, Polygon ZKEVM utilizes L1 calldata instead of blobs for DA. Transaction priority is determined via an auction-like system based on the effective gas price, allowing users greater control and predictability in transaction prioritization~\cite{polygonZKEVMFees}.

\subsection{zkSync Era}
Era explicitly separates L2 execution fees and L1 public data (pubdata) fees, giving users the flexibility to specify the maximum pubdata byte costs directly. Fees are locked upon batch sealing rather than immediate inclusion, offering potential refunds for unused resources, enhancing fairness and transparency. Users explicitly set their maximum fee limits, protecting them from unexpected volatility in L1 data posting costs. Transaction priority primarily relies on competitive L2 gas pricing, incentivizing efficiency and predictability~\cite{zkSyncEraFees}.

\subsection{Scroll}
Scroll's fees mirror Ethereum-equivalent execution costs. The L1 DA fees employ a deterministic oracle calculation factoring transaction size, the current L1 base fee, and an additional configurable overhead and scalar factor for risk management. Fees become fixed at transaction inclusion, transferring the risk of subsequent L1 gas price volatility entirely onto the sequencer. This provides users with predictable fees but places greater financial responsibility on the sequencer~\cite{scrollFees}.

\subsection{Optimism}
Optimism utilizes a fee model comprising L2 execution (following the EIP-1559 fee model) and an L1 DA posting fee that incorporates compressed transaction size along with configurable overhead and scalar factors. Oracle updates per block dynamically adjust these fees, locking in fees at the moment of inclusion to manage volatility effectively. Priority transactions follow the EIP-1559 tip mechanism, enabling users to explicitly bid for faster inclusion through increased priority fees. This dual-component system balances cost efficiency with operational simplicity~\cite{optimismFees}.

\subsection{Arbitrum}
Arbitrum's fee calculation combines L2 execution fees (also utilizing an EIP-1559-like mechanism) with the costs of posting compressed DA to L1, using individual transaction compression for accurate approximation. A dynamic oracle continuously adjusts fee parameters to stabilize pricing and recover actual L1 posting costs efficiently. Arbitrum introduces the TimeBoost mechanism~\cite{Mamageishvili2023BuyingTL}, a unique priority fee system where additional fees directly translate into transaction timestamp adjustments, allowing transactions to gain up to a $0.5$-second priority advantage. This innovative method mitigates traditional priority gas auction inefficiencies by converting priority fees directly into measurable temporal advantages~\cite{arbitrumDocs}. Further, Arbitrum enforces a gas speed limit of a target of 7M gas per second. Finally, the batch poster is designed to try and not overpay L1 gas, and keep backlog constrained (transactions posted to L2 and not yet to L1). Hence, if the backlog increases in size then the blob price will spike in the L2.

\begin{figure*}[t]
  \centering
  \begin{subfigure}[b]{0.45\textwidth}
    \centering
    \includegraphics[width=\linewidth]{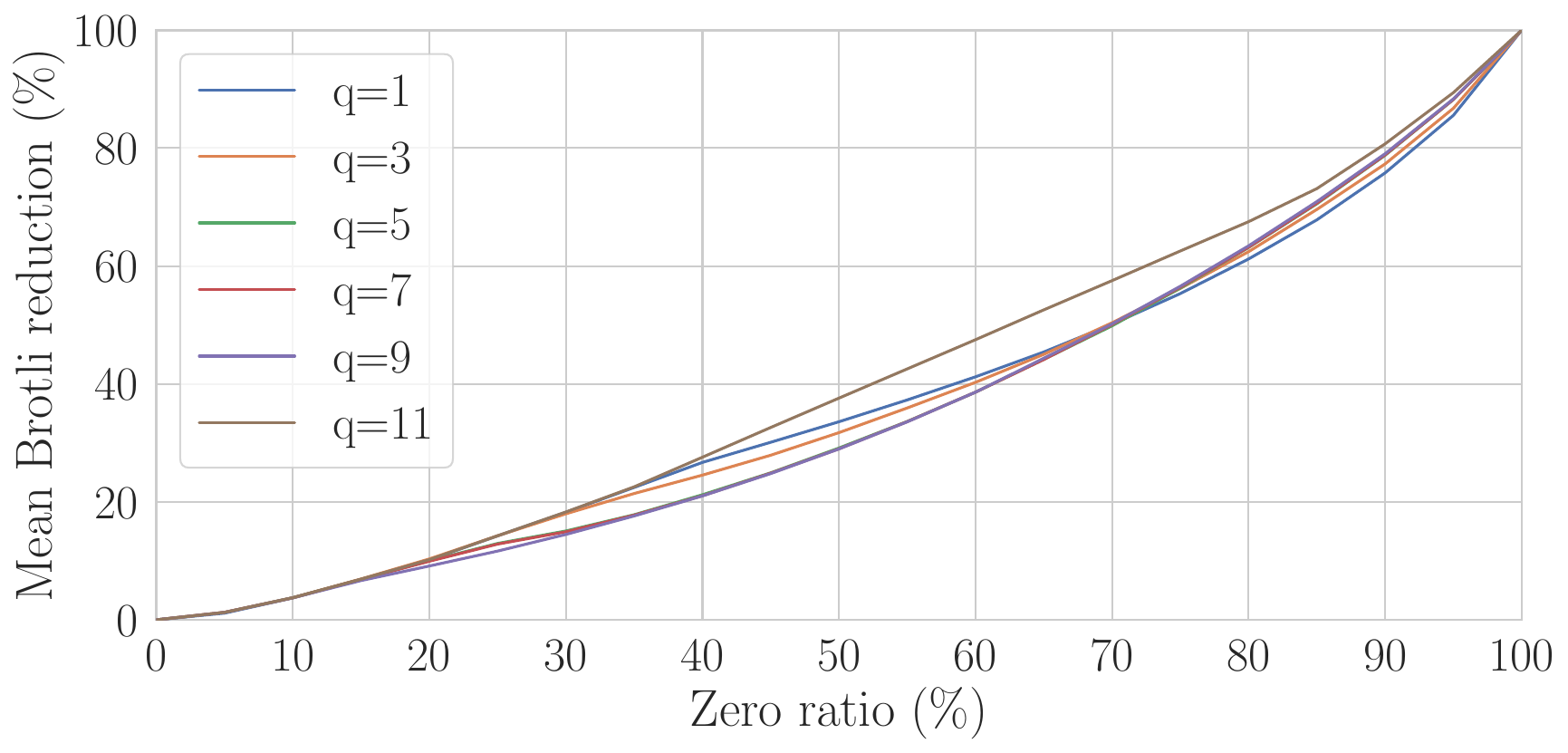}
    \caption{Brotli reduction vs zero ratio}
    \label{fig:compression_quality}
  \end{subfigure}
  \hfill
  \begin{subfigure}[b]{0.45\textwidth}
    \centering
    \includegraphics[width=\linewidth]{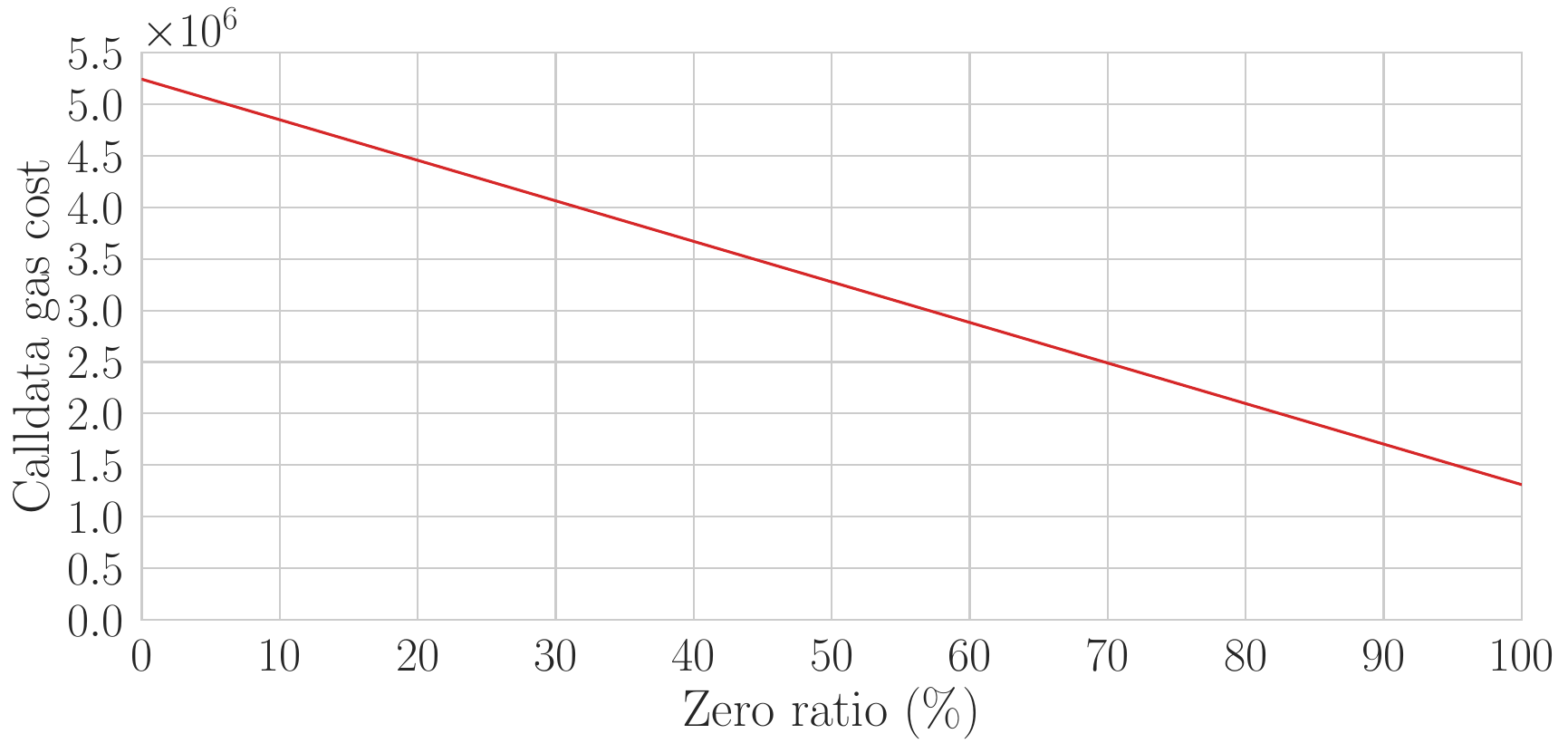}
    \caption{Calldata gas cost vs zero ratio}
    \label{fig:gas_quality}
  \end{subfigure}
  \caption{Comparison of Compression and Gas-cost across zero ratios for various Brotli qualities.}
  \label{fig:quality_comparison}
\end{figure*}

\begin{figure*}[t]
  \centering
  \begin{subfigure}[b]{0.45\textwidth}
    \centering
    \includegraphics[width=\linewidth]{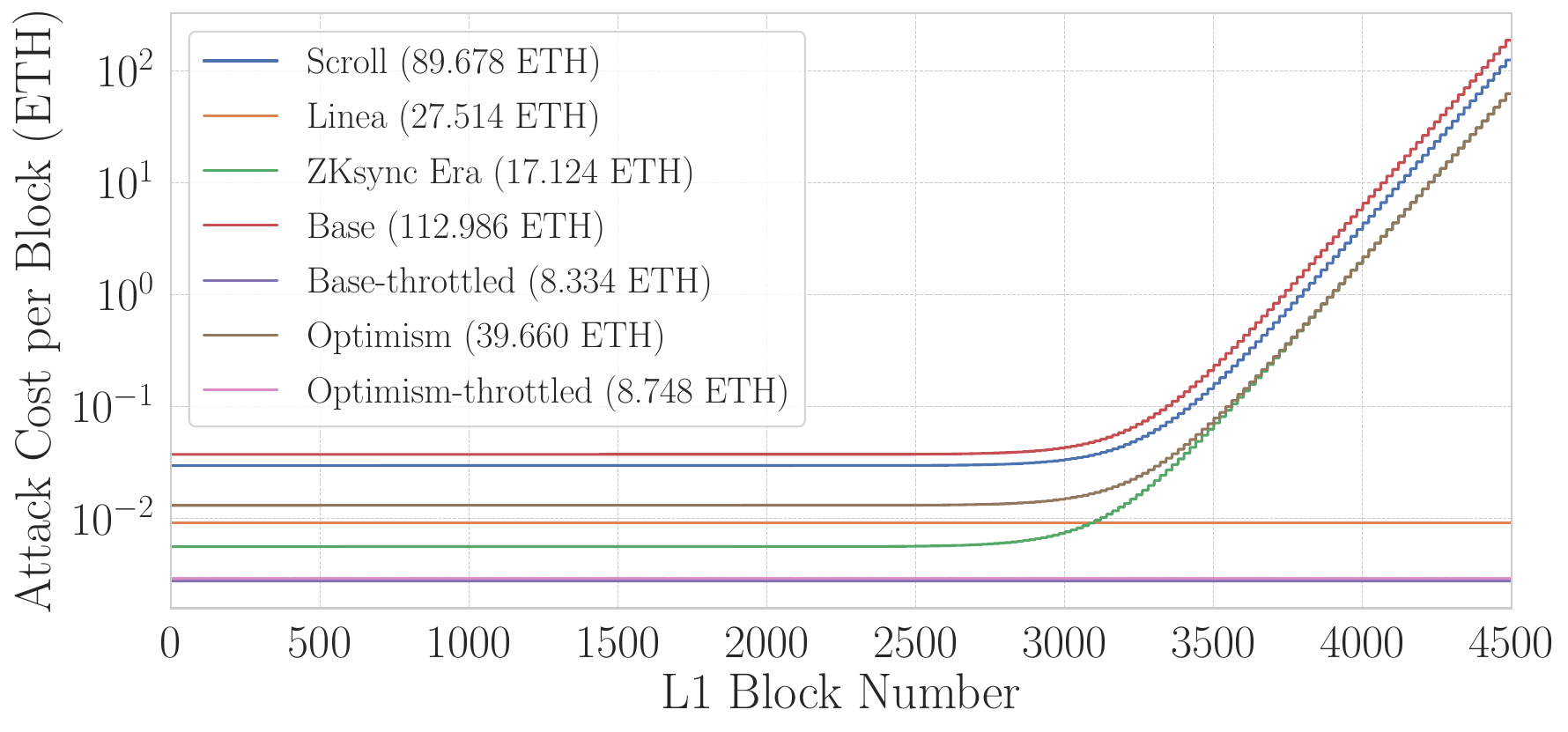}
    \caption{DA DoS Attack. Number in the legend are the accumulated costs for 3,000 blocks.}
    \label{fig:dados_prolonged}
  \end{subfigure}
  \hfill
  \begin{subfigure}[b]{0.45\textwidth}
    \centering
    \includegraphics[width=\linewidth]{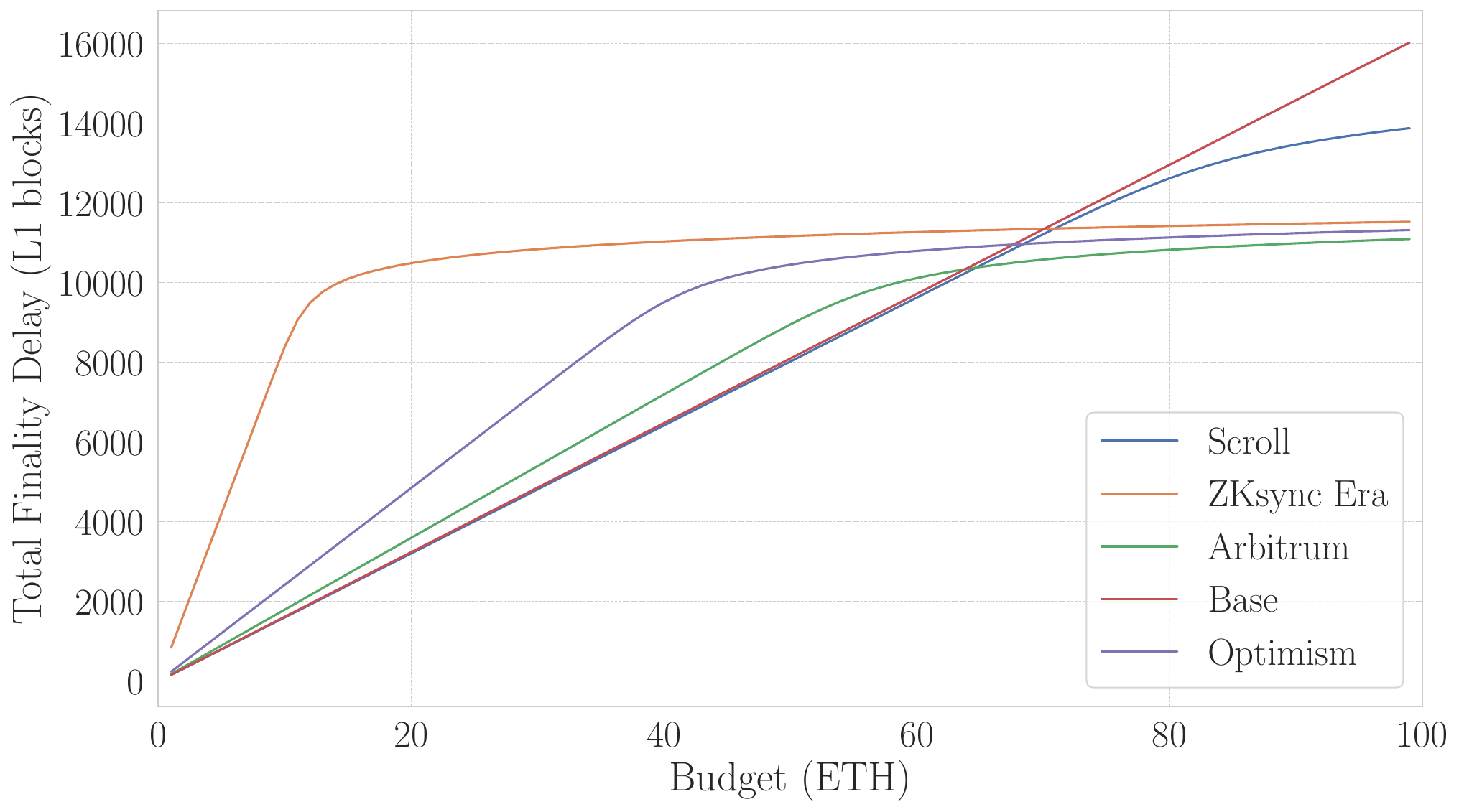}
    \caption{Amplified Finality Delay Attack.}
    \label{fig:l2da_prolonged}
  \end{subfigure}
  \caption{Impact of DA DoS and Amplified Finality Delay Attack when the rollups fetch L1 blob prices with a delay of 20 L1 blocks.}
\end{figure*}

\section{Compression Analysis}
\label{app:compression}

Ethereum's gas metering mechanism after the Pectra upgrade that includes EIP-7623, charges~10 gas units for every zero byte submitted in calldata and 40 gas units for every non-zero byte. Consequently, submitting predominantly zero bytes can significantly reduce calldata costs. However, rollups typically apply compression algorithms before posting data in blobs to Layer 1, which undermines strategies based solely on submitting zeros. Specifically, a blob filled entirely with zero bytes would be highly compressible.

In this paper, we adopt a conservative approach by using entirely random and hard-to-compress data for our analyses. Nevertheless, we conducted additional experiments using the Brotli algorithm, which is widely utilized by rollups, to evaluate how effective an attack could be if rollups employ suboptimal compression settings. For each experiment, we generated random byte sequences~100 times and reported the mean compression and cost values.

Figure~\ref{fig:compression_quality} illustrates the compression efficiency of Brotli relative to the percentage of zeros in the input data. Figure~\ref{fig:gas_quality} shows the corresponding calldata costs required to fully occupy one blob, based on varying zero-byte ratios.

\section{DA-Saturation Economic Attack: Scroll PoC}
\label{app:scroll-poc}

This appendix presents the details of a proof-of-concept that exploits all three attack vectors on Scroll, with a focus on the economic damage attack. 
For this attack, we use the pre-Pectra calldata cost, as at the time of writing, Scroll had not yet upgraded to the Pectra upgrade pricing.
Scroll has already patched the economic-damage vulnerability.

To evaluate the practical impact of these attacks, we deploy them on Scroll's devnet. Each crafted DA-heavy, compute-light transaction incurs:

\begin{itemize}
    \item Intrinsic Gas: $21{,}000$
    \item Calldata Gas: $2{,}080{,}064$ ($128$ KB of random, incompressible data)
    \item L1 DA fee: $52{,}428$ gas (using Scroll’s blob‐fee scalar at the time)
\end{itemize}

Our simulation shows this attack is feasible at a very low cost. We estimate the resources needed to block all L2 transactions for one hour ($1{,}200$ Scroll L2 blocks at the time). At a conservative L2 gas price of $0.25$ Gwei (to guarantee sequencing priority) and an ETH price of USD $2{,}500$, the one-hour costs are:

\begin{itemize}
  \item Intrinsic gas cost: $0.0063$ ETH (\$$15.75$)  
  \item Calldata gas cost: $0.6240$ ETH (\$$1{,}560$)  
  \item Blob fee cost: $0.0052$ ETH (\$$13.00$)  
\end{itemize}

Total: \$$1{,}588.75$ for one hour of full network disruption (\$$1.32$ per batch). This is negligible compared to the potential economic harm-manipulating oracles, delaying liquidations, or blocking critical transactions during volatility. At the time, Scroll’s mainnet commit transaction ($849{,}900$ gas at $5$ Gwei) cost $0.0067$ ETH (\$$10.62$) in L1 fees. Thus, for a $100$\%-fill batch, the protocol's outlay is \$$10.63$, while it collects only \$$1.32$ in fees, a net loss of \$$9.31$ per batch, or \$$11{,}172$ per hour.

To demonstrate replicability, we developed a Bash script that automates the attack:
\begin{itemize}
  \item \textbf{Random Calldata Generation:} Produce incompressible $128$ KB payloads.  
  \item \textbf{Transaction Pacing:} Dispatch one transaction per L2 block, matching Scroll’s block time.  
  \item \textbf{Asynchronous Submission:} Use multiple accounts to flood the mempool without awaiting confirmations.  
\end{itemize}

\subsection{Replication Environment}

Using the official Scroll SDK (\texttt{github.com/scroll-tech/scroll-sdk}), we configured a local testnet mirroring production, with details below.

\begin{enumerate}[leftmargin=*,nosep]
  \item \textbf{Local testnet:} Scroll devnet with default parameters.  
  \item \textbf{Node config tweaks:}
\begin{footnotesize}
\begin{verbatim}
"l2_config"."relayer_config"."sender_config" {
  "max_gas_price": 2000000000000000000,
  "max_blob_gas_price": 2000000000000000000,
  "confirmations": "0x0",
  "max_pending_blob_txs": 10000
}
\end{verbatim}
\end{footnotesize}
  This forces immediate batch commits regardless of L1 fees.
\end{enumerate}

\subsection{Observed Effects}

\begin{enumerate}[leftmargin=*,nosep]
  \item \textbf{Transaction Exclusion:} Honest transactions with equal or lower gas prices were routinely dropped, confirming a DoS condition.  
  \item \textbf{Fee undercharging:} Receipts showed the protocol consistently under‐collected fees: a $100$\%, fill batch cost \$$10.63$ in resources but yielded only \$$1.32$.  
  \item \textbf{Selection Nuances:} Scroll's L2Geth \texttt{pendingWithMax} pool selects txs pseudo-randomly by address, so a single‐address flood cannot fully starve other addresses. Nevertheless, the incompressible payloads still force the protocol to subsidize DA costs, inflicting net losses. Further, an attacker can use multiple addresses.
\end{enumerate}

\section{DA-Attack Impact when Blob Price is not Updated Immediately}
\label{app:delayed-updates}
In the attacks presented in Section~\ref{sec:da-saturation}, we assumed that the L1 blob price is retrieved immediately from L1. This assumption does not hold in practice, as blocks are not yet finalized. In this appendix, we analyze how the attacks of Sections~\ref{sec:da-dos} and~\ref{sec:da-finality} are affected when there is a delay in fetching the L1 blob price. Ethereum finality is typically achieved after 64 L2 blocks, however, some rollups may choose to wait for fewer blocks (commonly 5 or 20). In this appendix, we evaluate both attacks under the assumption of a 20-block delay in retrieving the L1 blob price. Depending on the rollup configuration, such an attack may either drain the rollup’s vault entirely or force batch posting to pause.

Figure~\ref{fig:dados_prolonged} illustrates the impact of fetching the L1 blob price with a 20-block delay for the DoS attack. In the legend, the cost is reported over $3{,}000$ blocks (approximately 10 hours). Notably, the cost remains sustainable up to $3{,}000$ blocks for all chains, indicating that the attack is even more damaging under these conditions.

Figure~\ref{fig:l2da_prolonged} depicts the impact of the amplified finality delay attack. For an expenditure of 10~ETH, the amplification of the attack relative to the L1 baseline ranges from $4.03\times$ (Scroll) to $21.11\times$ (zkSync Era). Importantly, if the attack persists over a long period, it could fully deplete the rollup vaults or force them to suspend batch posting. Increasing the delay to 64 blocks makes the attack even more severe.

%

\end{document}